\documentclass[final,5p,times,twocolumn]{elsarticle}

\usepackage{graphicx}
\usepackage{amssymb}

\usepackage{lineno}

\usepackage{amssymb}

\usepackage{amssymb}
\usepackage{amsmath}
\usepackage{siunitx}
\usepackage{tabularx}
\usepackage{algorithmic}

\usepackage{algorithm2e} 
\usepackage{amsthm}
\usepackage{graphicx}
\usepackage{lscape}
\usepackage{verbatim}
\usepackage{color,soul}
\usepackage{xcolor}
\usepackage{subfigure}
\usepackage{subcaption}
\usepackage{tabularx,ragged2e,booktabs,caption,array,multirow,multicol}
\usepackage{csquotes}
\usepackage{mathtools}
\usepackage{lineno}
\usepackage{amsthm}  
\usepackage{listings}
\usepackage{amssymb}
\usepackage{latexsym}
\usepackage{epsfig}
\usepackage{float}
\usepackage{xspace} 

\usepackage{MnSymbol,wasysym}

\journal{ }

\begin{document}

\begin{frontmatter}


\title{ An analysis of vaccine-related sentiments from development to deployment of COVID-19 vaccines}



\author[unsw]{Rohitash Chandra}
\ead{rohitash.chandra@unsw.edu.au} 
 
\author[iitgbio]{Jayesh Sonawane}   
\ead{piephi367@gmail.com}
\author[iit]{Janhavi Lande}    
\ead{l.janhavi@iitg.ac.in }

\author[unsw]{Cathy Yu} 
\ead{jiaxin.yu@student.unsw.edu.au}

\address[unsw]{Transitional Artificial Intelligence Research Group, School of Mathematics and Statistics, UNSW  Sydney, 
NSW 2006, Australia. }

\address[iitgbio]{Department of Biosciences and Bioengineering, IITG, India}

\address[cont]{*contributed equally to this work}

\begin{abstract}

Anti-vaccine sentiments have been well-known and reported throughout the history of viral outbreaks and vaccination programmes. The COVID-19 pandemic had fear and uncertainty about vaccines which has been well expressed on social media platforms such as Twitter. We analyse Twitter sentiments from the beginning of  the COVID-19 pandemic and study the public behaviour during the  planning, development and deployment of vaccines expressed in tweets worldwide  using a sentiment analysis framework via deep learning models. In this way, we provide visualisation and analysis of anti-vaccine sentiments over the course of the COVID-19 pandemic. Our results show a link between the number of tweets,  the number of cases, and the change in sentiment polarity scores during major waves of COVID-19 cases. We also found that the first half of the pandemic  had drastic changes in the sentiment polarity scores that later stabilised which  implies that the vaccine rollout had an impact on the nature of discussions on social media. 

\end{abstract}


\end{frontmatter}
 

\section{Introduction}
\label{S:1}

The \textit{coronavirus disease 2019} (COVID-19) is an infectious disease  caused by \textit{severe acute respiratory syndrome coronavirus 2} (SARS-CoV-2) \cite{Gorbalenya2020species,monteil2020inhibition,world2020coronavirus}    which became a global pandemic in 2020 \cite{cucinotta2020declares} and continues as a major disruption to social and economic activities worldwide. The influence of social media on pandemic-related public attitudes and behavioural developments is profound \cite{socialmedia}. During the ongoing COVID-19 pandemic, social media platforms such as Twitter and Facebook have been heavily used for timely information sharing and communication \cite{goel2020social,cinelli2020covid}. Such user-generated contents contribute to a spectrum of opinions ranging from official announcements to expression of individual beliefs, from credible health updates to the dissemination of rumours and misinformation \cite{tasnim2020impact,sharma2020coronavirus,mian2020coronavirus,apuke2021fake}. This has facilitated diverse public sentiments towards COVID-19 and its control strategies given  a wide range of topics such as racism, deaths, and economic losses \cite{abd2020top}. Some regional studies have identified overall positive sentiments initially \cite{barkur2020sentiment}, despite the polarity of sentiments demonstrated on certain pandemic-related topics such as  quarantine measures \cite{karami2020social}, mask-wearing \cite{kumar2020knowledge} and anti-vaccination \cite{burki2020online,puri2020social}.

In the history of viral infections, antivaccine activities   have been well known and reported during  outbreaks and vaccination programmes \cite{blume2006anti,hussain2018anti,berman2020anti,kata2010postmodern}, such as refusal of parents to vaccinate children in USA \cite{olpinski2012anti} and during an outbreak of   measles in 2019 in USA \cite{benecke2019anti}. In a related study,  Figueiredo et al. \cite{de2020mapping} mapped global trends in vaccine confidence   across 149 countries between 2015 and 2019 and estimated     that  confidence  in   vaccines fell in several Asian countries and  improved  in some of the  European Union member states. The study found  a  link  between religious  beliefs and vaccine updates  and reported that a link between religious beliefs and vaccine  uptake. Social media has served as a tool for dissemination of official information and connectivity during lockdowns \cite{goel2020social,cinelli2020covid}; however,  it has been a tool for anti-vaccine  activities and movements also known as "anti-vaxxers" \cite{megget2020even}. Fear and uncertainty, due to abrupt changes in lockdowns during COVID-19 had a huge effect on mental health which includes patients \cite{yao2020patients} and general population  \cite{pfefferbaum2020mental,cullen2020mental,world2020mental} along with children   \cite{liu2020mental}; it was highlighted that mental health disorders can increase the risk of infections, and  barriers in accessing timely health services. Anti-vaxxer movements are also based on notions that are built from conspiracy theories and pseudo-scientific viewpoints \cite{kata2010postmodern}; however, at times they come from the adverse nature of the official vaccine itself, such as the March 2020 Astrazena vaccine ban in 18  countries.

Recent progress in deep learning models has improved language models \cite{otter2020survey}. Recurrent neural networks (RNN) have been prominent for language translation \cite{liu2014recursive,lakew2018comparison} and sentiment analysis tasks \cite{
basiri2021abcdm,li2016text}. The long-short term memory  (LSTM) network \cite{hochreiter1997long} is a prominent RNN that has been a backbone of  several prominent language models \cite{otter2020survey}.  There has   been some progress in improving LSTM models further with attention mechanism \cite{vaswani2017attention} and Transformer models \cite{karita2019comparative}  that combine attention and other novel innovations in LSTM models. The Transformer model  has been prominent in developing pre-trained language  models such as bidirectional
encoder representations from transformers  (BERT) \cite{devlin2018bert} for masked language modelling.

The effect  of misinformation is becoming severe and hence  there have been discussions about criminalising misinformation on social media.  Mills and Sivela \cite{mills2021should} presented a discussion where the opposing notion to criminalising anti-vaccine activities was that  the right to freedom of   expression although there can be a restriction for certain cases, such as inciting lawless activities and violence, where  anti-vaccination misinformation was not seen such a case.  Johnson et al. \cite{johnson2020online} pointed out  that there is an online competition with  and against vaccination and studied  nearly
100 million individuals are partitioned into highly dynamic, interconnected clusters around the globe across  languages. The 
anti-vaccination clusters had smaller sizes but manage to become highly entangled with undecided
clusters, and the  study  predicted that 
anti-vaccination views  will dominate in a decade.  In the age of artificial  intelligence and social media analysis 
\cite{xue2020public,hung2020social,wang2020covid}, sentiment analysis could be seen as a way to understand public behaviour towards vaccines which can lay out a framework for policy development. Social media has been used as a tool for studying pandemics \cite{tang2018social} in the past which covers viral outbreaks such as measles \cite{mollema2015disease,kim2020psychology} and management of  H1NI viral outbreak   \cite{freberg2013managing}. However, there has not been done much study that looks at public sentiments in relation to how they have expressed their views regarding vaccinations. Sentiment analysis can provide an indication of how a person is reacting towards the vaccination process; e.g. if a tweet has the term COVID-19 and vaccination and a trained sentimental analyse model classify the tweet as ``fear" and ``pessimistic", then it would lean towards anti-vaccination. Hence, this way, common sentiment detection can guide an understanding of misinformation regarding vaccination.

   In this paper, we analyse the sentiments from the beginning of  the COVID-19 pandemic and study the behaviour during the  planning, development and deployment of vaccines expressed in tweets worldwide  using a sentiment analysis framework. We train the model using  the Senwave sentiment analysis dataset which features 10,000 tweets during COVID-19 with 10 sentiments labelled by 50 experts \cite{yang2020senwave}. Furthermore, we use the pre-trained BERT language model  for comparison.    We present sentiment analysis and compare selected countries such as Australia, Japan, India, Brazil and Indonesia. In our framework, we define a set of sentiments to detect anti-vaccine sentiments and provide longitudinal data analysis. We use the trained model to predict sentiments associated with the term vaccine from tweets worldwide  for about two years since the beginning of COVID-19. In this way, we provide an analysis of monthly anti-vaccine sentiments over the course of the COVID-19 pandemic.

The rest of the paper is organised as follows. Section 2 presents   a review of related work that examines language models and sentiment analysis during COVID-19. Section 3 presents the proposed methodology with Twitter data extraction details along with   a deep learning framework. Section 4 presents experiments and results. Section 5 provides a discussion and Section 6 concludes the paper with a discussion of future work. 

\section{Related Work} 

Topic modelling and sentiment analysis make some of  the major studies of social networks using language models typically powered by deep learning methods. 
 Xue et al. \cite{xue2020public}  used  topic modelling and sentiment analysis  for about 1.9 million tweets   related to COVID-19  during the early stages  and categorized them into ten themes. The  sentiment analysis showed that fear of the unknown nature of the coronavirus was dominant in the respective themes. Hung et al. \cite{hung2020social} presented 
  a social network analysis of COVID-19 sentiments based on tweets from the United States  to determine the social network of dominant topics and type of  sentiments with geographic analysis and found five prevalent themes which could clarify   public  response   and help officials. Wang et  al. \cite{wang2020covid} presented  sentiment and trend analysis of social media in China via a BERT-based model. Chakraborty et al. \cite{chakraborty2020sentiment} presented
 sentiment analysis of COVID-19 tweets via deep learning  with handles related to COVID-19 and World Health Organisation and found that the  Tweets have been unsuccessful in guiding people. Abd-Alrazaq et al.  \cite{abd2020top} presented a study to find the key concerns of Tweets during the COVID-19 pandemic and   identified 12 topics with  themes such as ``origin of the virus",  ``its impact on people",   and ``the economy". Other related work on sentiment analysis during COVID-19  focused on areas of managing  diabetes \cite{cignarelli2020diabetes}, where a change in sentiments expressed was shown when compared to pre-COVID-19. Furthermore, region-specific studies included community sentiment analysis in Australia \cite{zhou2020examination}, and nationwide sentiment analysis in Nepal \cite{pokharel2020twitter} during the early months where a majority of positive sentiments were expressed with elements of fear.  Moreover,   sentiment analysis  in the case of Spain examined  how social media and digital platforms created an impact \cite{de2020sentiment}. Furthermore,  sentiment analysis was used to study the effect of  nationwide lockdown due to the COVID-19 outbreak in India with a majority positive response for early lockdowns   \cite{barkur2020sentiment}. A study of  European   cross-language sentiment analysis in the early  COVID-19 pandemic  separated the results by country of origin, and temporal development \cite{kruspe2020cross}
 

\section{Methodology}

 \subsection{Data Extraction from Twitter} 
 
  The COVID-19 tweets dataset \cite{dataset1} retrieved from \textit{IEEE DataPort} captures the tweet identifiers (IDs) and their associated sentiment scores for daily COVID-19-related feeds  in English using over 90 commonly used keywords and hashtags. The full details (i.e exact tweet content) of the raw tweet IDs were then retrieved and processed using tweet IDs through a software extension known as \textit{Hydrator} \cite{hydrator}, in compliance with the Twitter policy which prohibits the direct publication of tweets as openly available datasets.

  We  separated the  tweets from the global dataset for the selected countries in our study selected  which included  Australia, Japan, India, Brazil and Indonesia.   The dataset is published in Kaggle \cite{janhavi-data}.
  We note that it is up to users to decide if their geographical location is known while tweeting and hence we have a limited number of tweets with geolocation. 
 
 

The special phrases, emotion symbols (emoji's) and abbreviations that  were used in tweets need to be processed and translated into known English words as shown in  Table \ref{fig:framework}. The original dataset was restricted to tweets in English only.  
 
\begin{table*}[htbp!]
    \small
    \begin{tabularx}{1\textwidth}{l l} 
    \hline
    Tweet language usage & Standardised word usage  \\ 
    \hline
    
    omg & oh my god  \\
    tbh & to be honest  \\
    rt & retweet\\
    dm & direct message\\
    socialdistance & social distance\\
    fwiw & for what it’s worth  \\
    covid19vax & COVID-19 vaccine\\
     \smiley{} or \blacksmiley{} & smile     \\
     \frownie{} & sad    \\
    \hline
    \end{tabularx}
    \smallskip
    \caption{Changing tweet language usage and emojis to standardised language usages with semantic meaning.}
    \label{tab:process}
\end{table*}

 \subsection{BERT-based model}
 
     RNNs distinguishes   from feedforward neural networks due to feedback (recurrent) connections \cite{elman_Zipser1988,Elman_1990,Werbos_1990}. Moreover, RNNs   feature  a  context memory  layer that is used to implement the recurrence in order to compute  future state outputs.  The earlier architectures are also known as   simple RNNs   which have been prominent  for modelling temporal sequences \cite{ChandraTNNLS2015}  and dynamical systems \cite{Omlin_etal1996,Omlin_Giles1992}.   A major limitation in simple RNNs was the difficulty to train using   backpropagation through time which extends the  backpropagation algorithm  \cite{Werbos_1990}. The major issue was  difficulty in learning long-term dependencies given  vanishing/exploding gradients  \cite{Hochreiter_1998}.  LSTM  \cite{hochreiter1997long} networks addressed the limitations of using memory cells and gates for    better capabilities in remembering long-term dependencies. Bi-directional LSTM networks \cite{graves2005} make use of only the previous context state for determining the next  states which enables them to process information in two directions and built on the ideas from  bidirectional RNNs  \cite{schuster1997}.  Bi-directional LSTM networks   were initially designed for word-embedding and  have been used in several other natural language  processing problems \cite{graves2005,Fan2014TTSSW,graves2013hybrid}, and later been extended as the Transformer and BERT models.   BERT-based  models have been used for a wide range of sentiment analysis-related tasks such as modelling US 2020 Presidential elections \cite{chandra2021biden}, translation analysis  of Bhagavad Gita which is a Hindu philosophical and sacred religious text 
\cite{ChandraKulkarni2022}, and  COVID-19 sentimental analysis with a study of the first wave of the effect of COVID-19 in India \cite{chandra2021covidsentiment}. BERT has also been used for topic modelling that includes a study of common topics present in the Hindu texts that included the Bhagavad Gita and Upanishads \cite{chandra2022artificial}, and COVID-19 Twitter-based  topic modelling for three different waves in India \cite{lande2023deep}.

     
   
     
 \subsection{Framework}
 
  Our overall goal is to obtain a visualisation and provide analysis of tweets relating to vaccination during, planning, developing and deployment of vaccines in the COVID-19 pandemic. We implement sentiment analysis using a BERT-based model  to understand the nature of the tweets, in terms of overall negative or positive sentiments. We note that anti-vaxxer tweets on  their own do not explicitly state that they are anti-vaxxers. They express their ideals in a way that promotes fear and uncertainty associated with vaccination and at times the expression would be seen as very professional while being subtle in the message against vaccination. Hence, it is not straightforward to determine  the anti-vaxxer tweets. We hence present a framework where  sentiment analysis   is used as a methodology to detect anti-vaxxer tweets. 
  
\begin{figure*}[htbp!]

\hbox{\hspace{3em}\includegraphics[width=17cm]{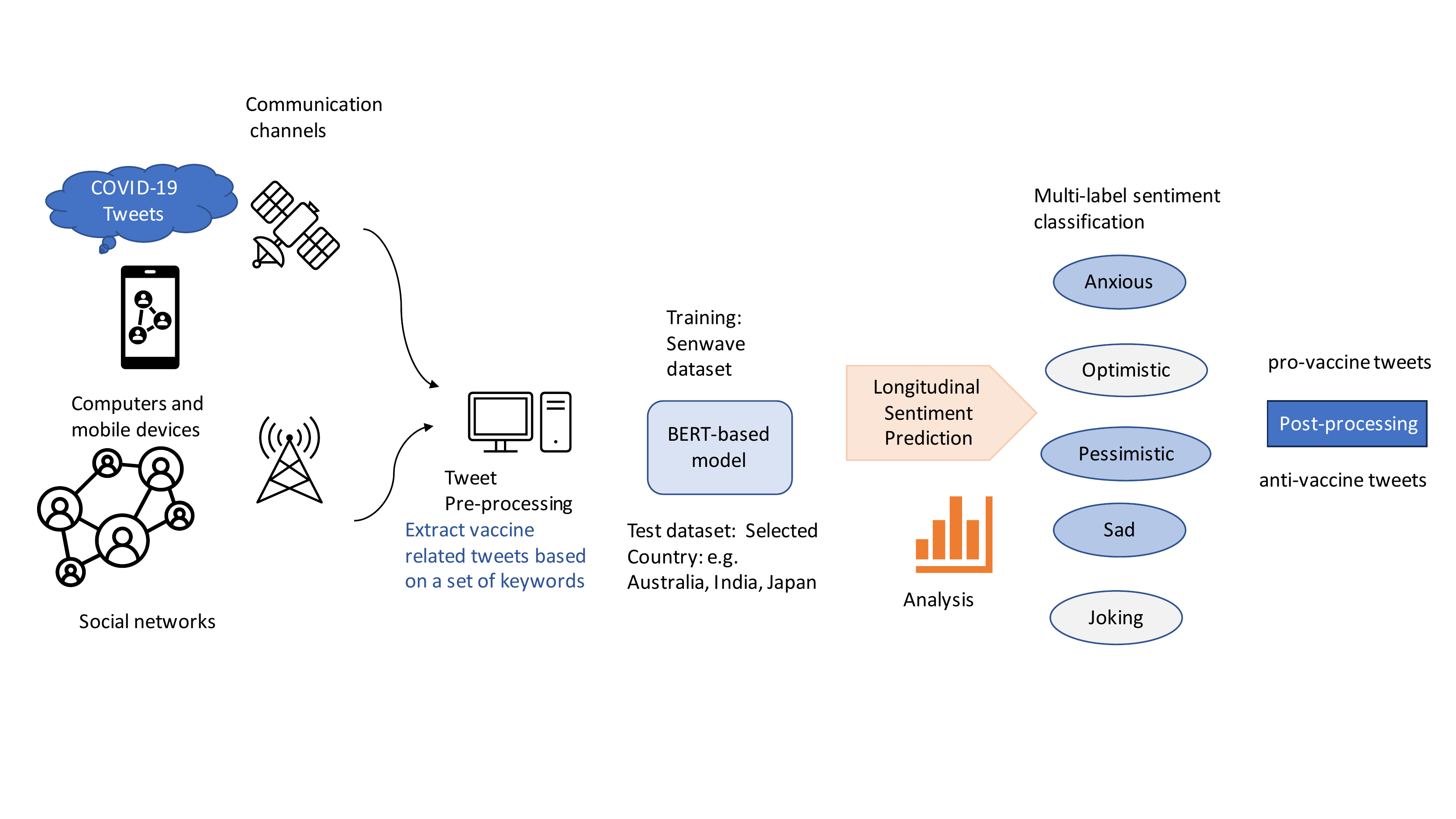}}
\caption{Framework for using sentiment analysis for detecting anti-vaxxer tweets for COVID-19 via deep learning models.}
\label{fig:framework}
\end{figure*}

  We propose BERT-based sentiment analysis framework shown in Figure \ref{fig:framework} to support our investigation of anti-vaccine sentiment movements during the planning, development and distribution phases of COVID-19 vaccinations. 
 
   Figure \ref{fig:framework} presents the framework  for detecting anti-vaxxer tweets using sentiment analysis with  the following steps; 1.) Tweet extraction using application software and  pre-processing, 2.) word embedding using BERT model from Tweets, 3.)  model development and training via BERT, and 4.) post-processing of sentiments to categorise prop-vaccine, anti-vaccine and neutral vaccine-related tweets  during COVID-19.  The  model training features  multi-label classification where a tweet  can feature more than one sentiment at a time, i.e.  a tweet can be optimistic and anxious. Hence, it is not  simple to determine negative and positive sentiments when we look at more than one outcome. A multi-label classification model utilises training sets where each instance can have a set of labels attached to it, and the goal is to assign multiple labels to any unseen instances for prediction. We note there is a distinguishing difference between a multi-class classification problem and a multi-label classification problem. In a multi-class classification task, all classes are mutually exclusive and each instance can only be assigned with one label, whereas in the above-defined COVID-19 sentiment multi-label classification task, each instance can be assigned with multiple labels given a separate classification mechanism will be triggered for the classification of each label.

The SenWave is a unique multi-labelled COVID-19 sentiment dataset that was developed by researchers in 2020 \cite{yang2020senwave}. Over 10,000 COVID-19 related tweets in the English language were manually assigned one or more emotional labels including ‘optimistic’, surprise', ‘thankful’, ‘empathetic’, ‘pessimistic’, ‘anxious’, ‘sad’, ‘annoyed’, ‘denial’, ‘official’, and ‘joking’ which forms the basis for language model training. In this study, the sentiment official' is discarded. The Senwave dataset is different from most traditional sentiment analysis datasets since they mostly have  sentiment scores associated with ‘positive’, ‘negative’ and ‘neutral’ sentiments. Furthermore, the Senwave is more applicable to our study since the dataset is from COVID-19 tweets.

We fine-tune the pre-trained BERT model  using  Senwave COVID-19 dataset.  Before training, the tweets are pre-processed (using Table 1) and each   word in the tweet is given   a corresponding world  embedding vector (as shown in Figure \ref{fig:framework}). The word  vector for each tweet features SenWave hand-labelled sentiments as outcomes   which are  used to train   the BERT-based models  as shown in Figure \ref{fig:framework}.   Once we get the tweet sentiment classified by the  model, we make some assumptions and devise a strategy to determine if the tweet is an anti-vaxxer, prop-vaccination or neutral using a vaccine score that we compute by using an average of the scores. Figure \ref{fig:framework} shows further details about the post-processing of the sentiments classified by the BERT-based model.  

 We also compute the sentiment polarity  score using a textual data processing library in Python known as  \textit{TextBlob}.
  The score is given  in a range (-1 to 1), where a positive score indicates positive sentiment toward the statement and a negative score indicates  negative sentiment.

\begin{figure}[htbp!]
\centering
\subfigure[Positive sentiments combinations]{
   \includegraphics[height=5cm]{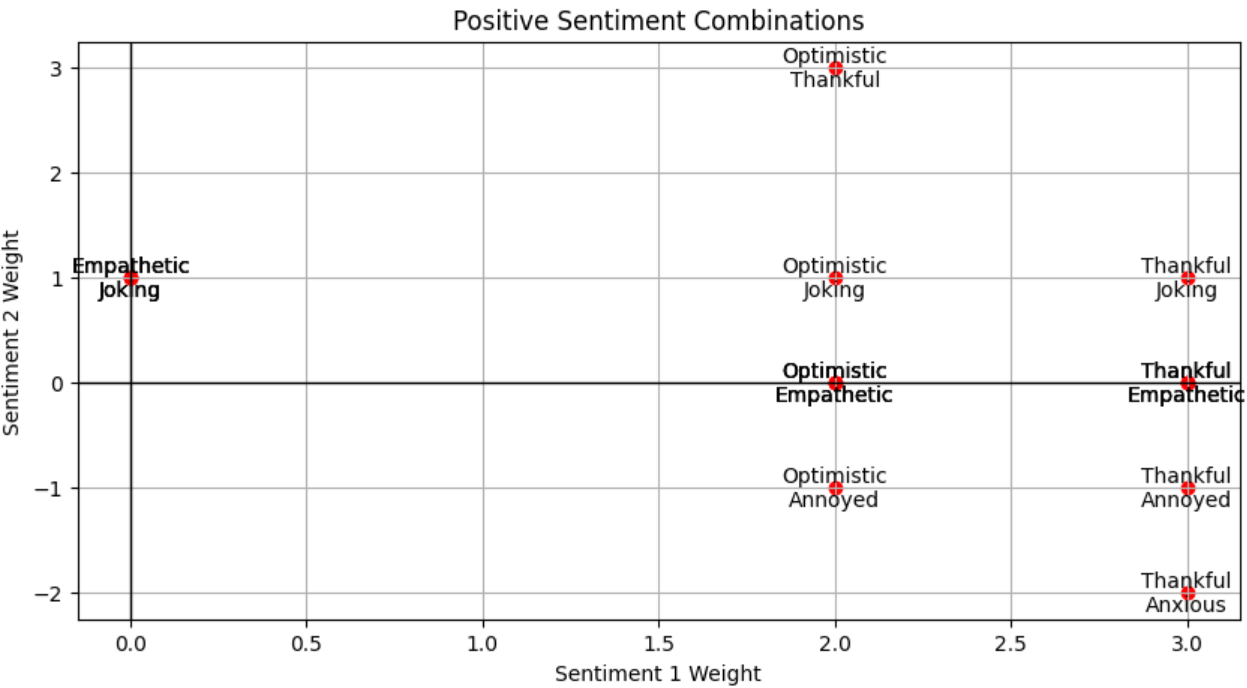}%
}
\subfigure[Neutral sentiments combinations]{
   \includegraphics[height=5cm]{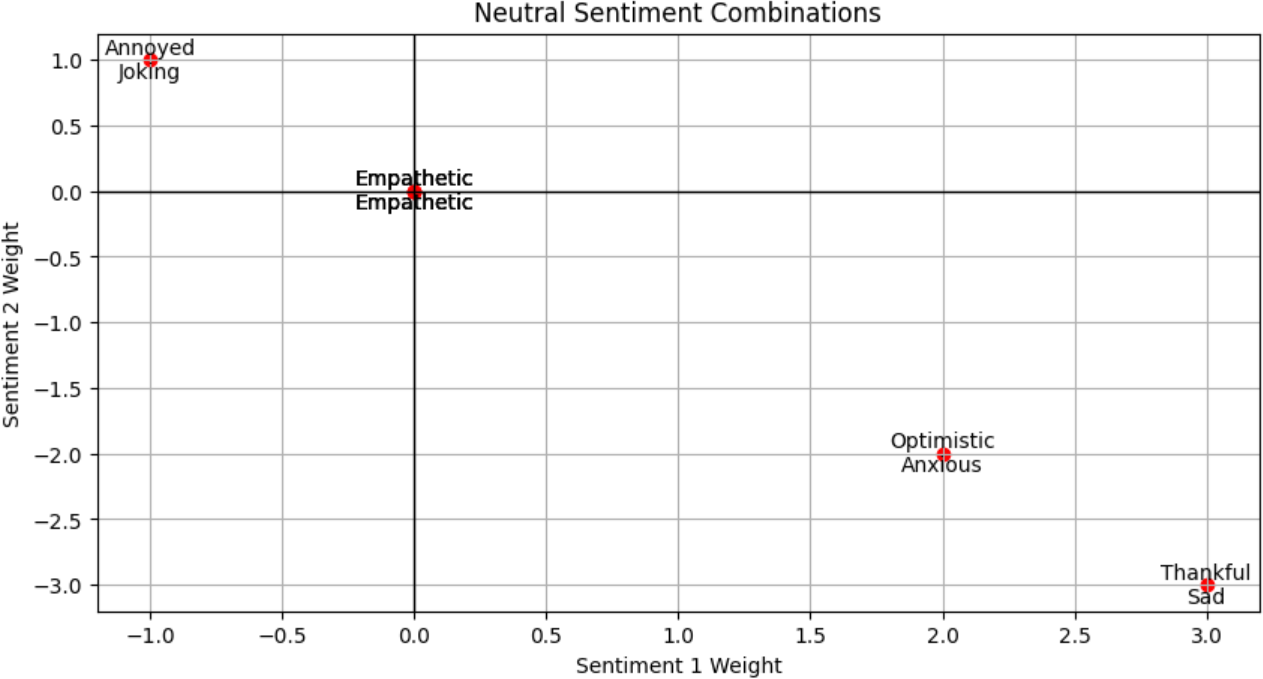}%
}
\subfigure[Negative sentiments combinations]{
   \includegraphics[height=5cm]{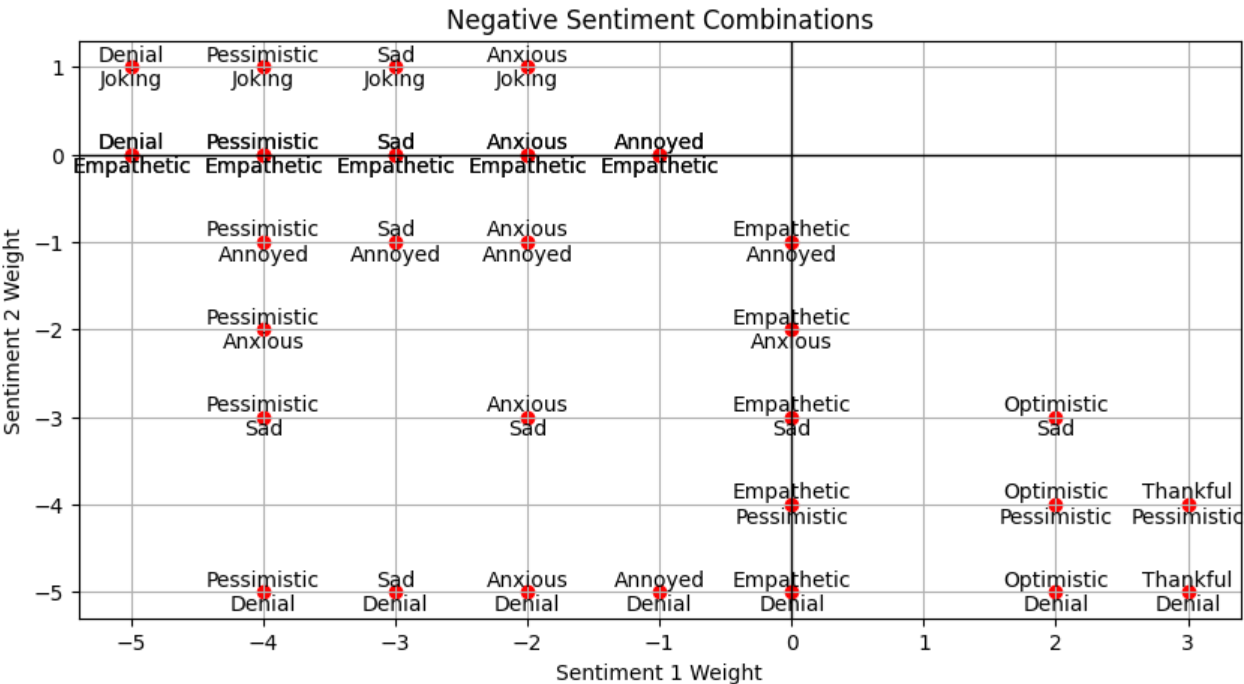}%
}
 \caption{Combination of two sentiments labels for  vaccine polarity score where we simply add weights of Sentiment 1 and Sentiment 2. }
\label{fig:combinations}
\end{figure}

Finally, we assign weights to each sentiment to calculate our own vaccine polarity score as shown in Table  \ref{tab:score}. In the case of two sentiment labels, Figure \ref{fig:combinations}  presents combination of  the sentiments labels to calculate the  vaccine polarity score where we simply add weights of Sentiment 1 and Sentiment 2. This is used for post-processing as shown in Figure 1.

\begin{table}[htbp!]
    \small
     \centering
    \begin{tabular}{p{4cm}p{2cm}}
    \hline
    \textbf{Sentiments} &  \textbf{Sentiment weight} \\ 
    \hline
     Optimistic &  2 \\%
     Thankful & 3\\%
     Empathetic & 0\\%
     Pessimistic & -4\\%
     Anxious & -2\\%
     Sad & -3\\%
     Annoyed & -1\\%
     Denial & -5\\%
     Official report & 0\\%
     Surprise & 0 \\%
     Joking & 1\\%
     \hline 
    \end{tabular}
    \smallskip
    \caption{Weight assigned  for each sentiment to calculate vaccine polarity score. }
    \label{tab:score}
\end{table}

\subsection{Technical Setup} 

We adopt the technical setup of our framework  from Chandra and Krishna \cite{chandra2021covidsentiment}  where the SenWave dataset and BERT-based model were used for sentiment analysis in India during the rise of COVID-19. The  BERT-based model  features  11 outputs in the output layer.  We aim to capture the interrelationships among those sentiment labels to better understand the overall semantic classification of tweets. After model training, we present country-wise tweets and record the sentiment prediction by the model for longitudinal analysis, visualisation, and post-analysis. 

\section{Results}  

\subsection{Data Analysis}  
  
\begin{figure}[htbp!]
 \centering
 \includegraphics[width=9cm]{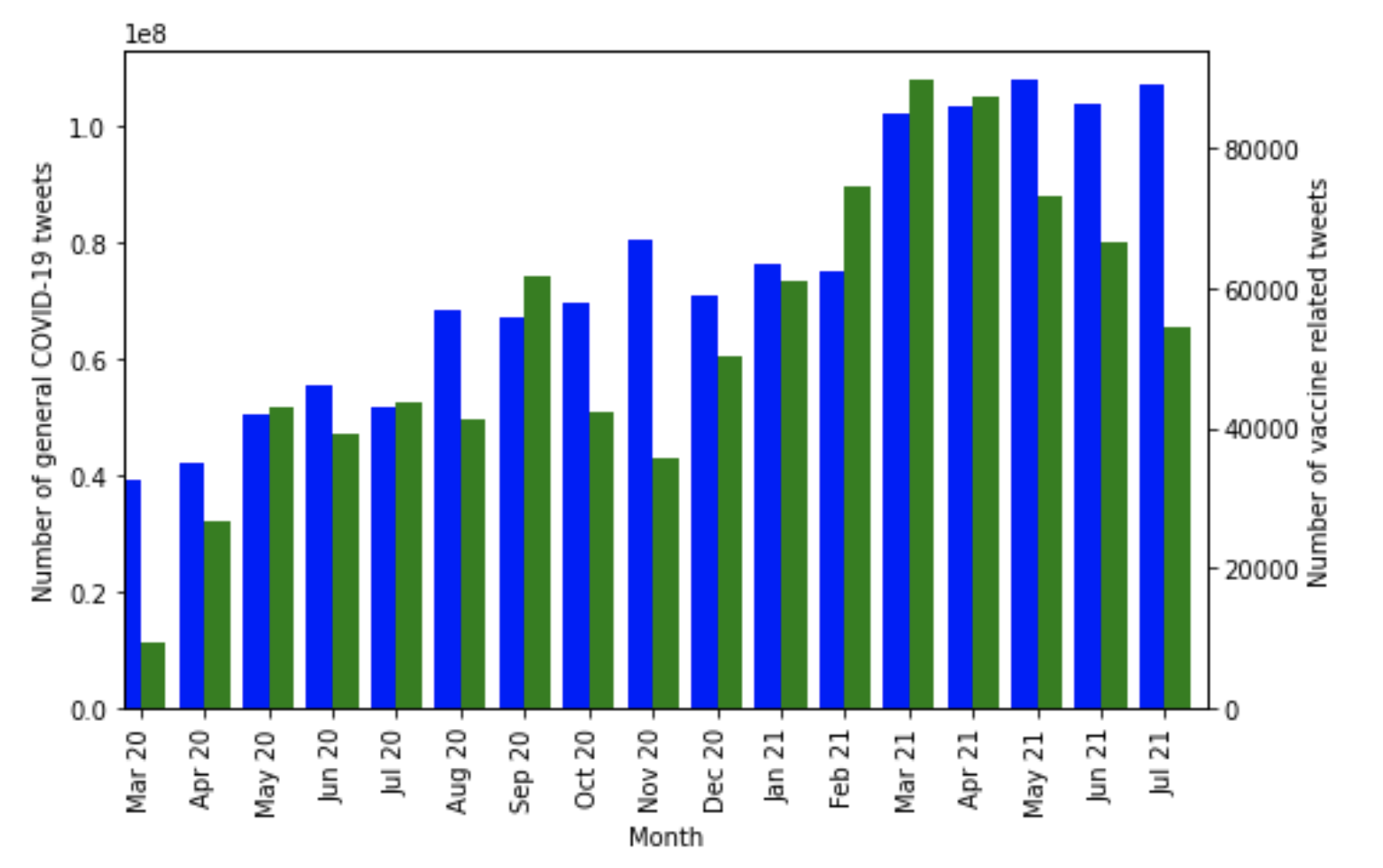}
 \caption{Comparison between general COVID-19 tweet counts (blue) and COVID-19 vaccine-related tweets (green)  worldwide   from March 2020 to July 2021. }
 \label{fig:count}
 \end{figure}

We first present data analysis for extracted tweets related to COVID-19. 
Figure \ref{fig:count} illustrates a comparison between the volume of vaccine-related tweets and general COVID-19-related tweets at the begging of the pandemic (March - July 2020).   We notice that there are fewer vaccine-related tweets during the beginning of the pandemic (March and April  2020) when compared to the months that follow (May, June and July 2020). Afterwards, there is a decline and then a rise in vaccine-related tweets. These trends would align with discussions by mainstream media, vaccination rate and related factors that give rise to such tweets.

\begin{figure}[htbp!]
\centering 
\subfigure[ Bigram ]{
    \includegraphics[height=4cm]{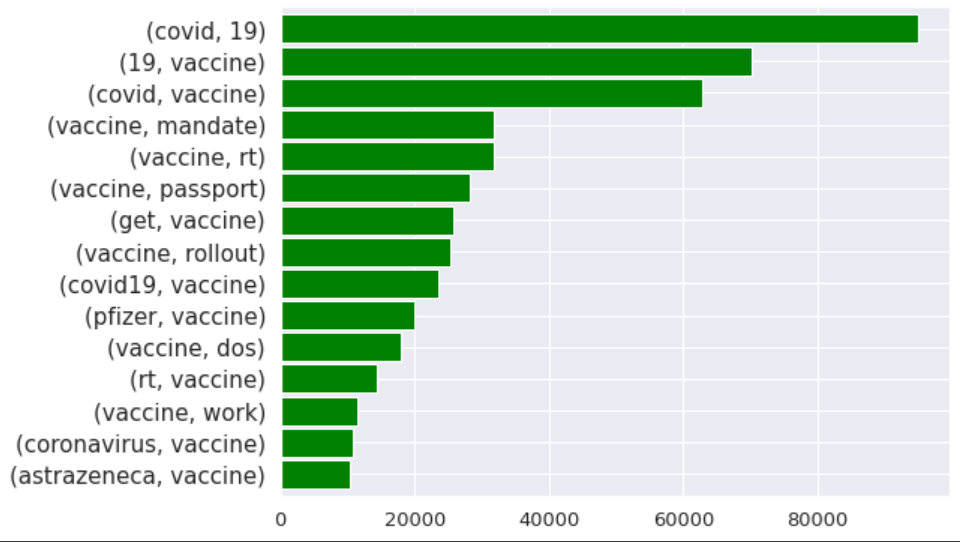}%
}
\subfigure[ Trigram]{
    \includegraphics[height=4cm]{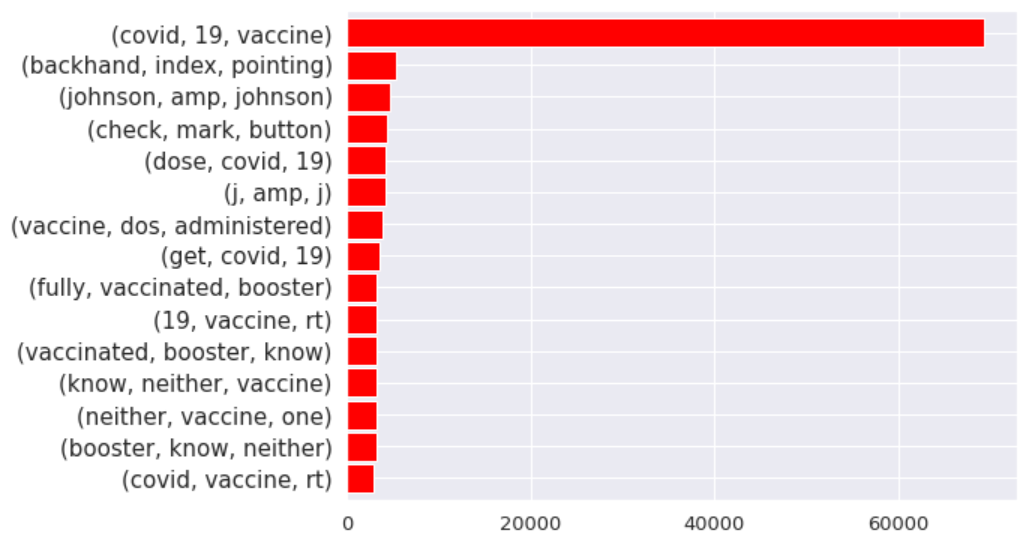}%
}
\caption{ Bi-gram and tri-gram analysis of vaccine-related tweets posted between March 2020 and July 2021 worldwide.}
\label{fig:n-gram} 
\end{figure}

N-gram \cite{brown1992class} analysis is a method frequently used in NLP to capture the inter-dependencies among sequences of words of length $N$ in textual data. We conducted a word-level bi-gram and tri-gram analysis  from March 2020 - July 2021 worldwide to find occurrences of consecutive words in the selected tweet dataset as shown in Figure \ref{fig:n-gram}. We observe that in the bi-grams, the phrase ‘covid19 vaccine’ has the highest frequency of appearance, closely followed by ‘vaccine force’ and ‘waning immunity’. In the tri-grams, the three-worded phrase ‘fight covid19 smile’ (the smiling emoji has been converted into the word ‘smile’) is most commonly used, followed by ‘covid19 positive test’ and ‘positive antibody test’.

\subsection{Model Prediction}

\begin{figure}[htbp!]
\centering
\includegraphics[width=10cm]{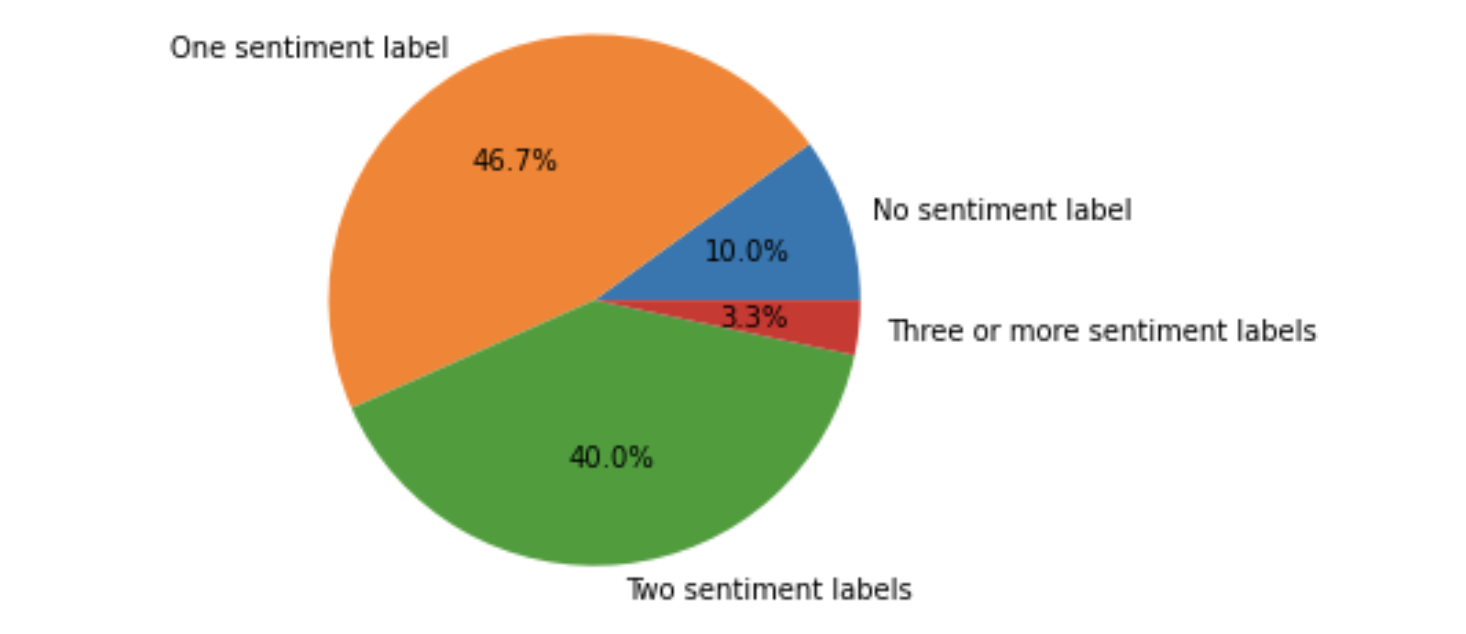} 
\caption{Percentage of COVID-19 vaccine-related tweets with different numbers of sentiment labels predicted by the LSTM model.}
\label{fig:pie}
\end{figure}

Next, we present the multi-label sentiment classification results using the proposed framework for worldwide vaccine-based COVID-19 tweets. Figure \ref{fig:pie} presents a summary of the number of sentiment labels assigned to all 850,000 vaccine-related tweet instances. As shown, approximately 10\% of the input tweets were assigned with no sentiment labels, 46.7\% with one unique sentiment label, 40\% with two sentiment labels, and 3.3\% with three or more sentiment labels.

\subsection{Sentiment analysis by country}

\begin{figure}[htbp!]

\centering 
    \includegraphics[height=5cm]{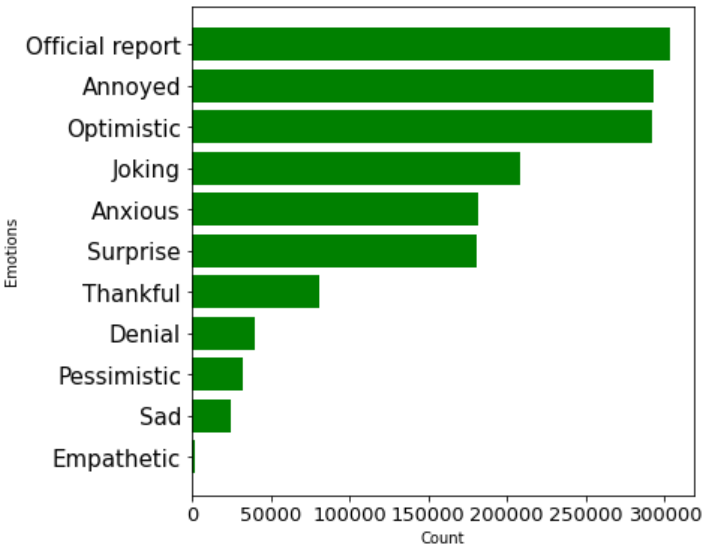}%
\caption{Number of vaccine-related tweets for each sentiment}
\label{fig:keywords} 
\end{figure}

\begin{figure}[htbp!]
 \centering
    \includegraphics[height=4cm]{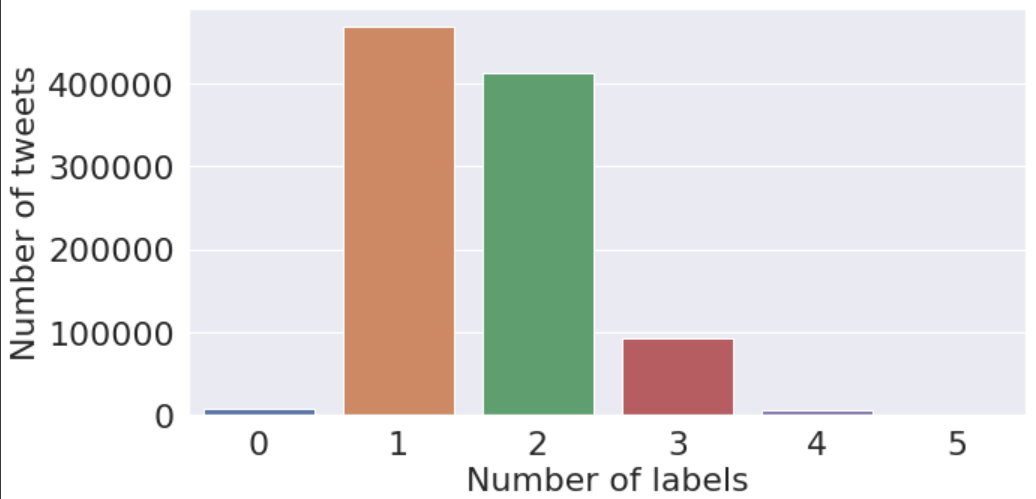}%
\caption{ Analysis plot of the number of tweets corresponding to the number of sentiments. }
\label{fig:labels} 
\end{figure}

Finally, we  present the longitudinal analysis of sentiments in COVID-19 related tweets over the time period from March 2020 to February 2022 based on the 11 sentiment labels. Figure \ref{fig:keywords} gives an idea about how tweets are distributed according to their sentiments. From this plot, we can observe that 300,000 tweets correspond to the sentiment "Official report", Followed by "Annoyed" and "Optimistic". This shows the sentiments that have dominated among Twitter users during the COVID-19 pandemic.   Figure \ref{fig:labels}   suggests that more than 400,000 tweets correspond to only one sentiment followed by only two sentiments.



\begin{figure}[htbp!]
\centering  
    \includegraphics[height=4cm]{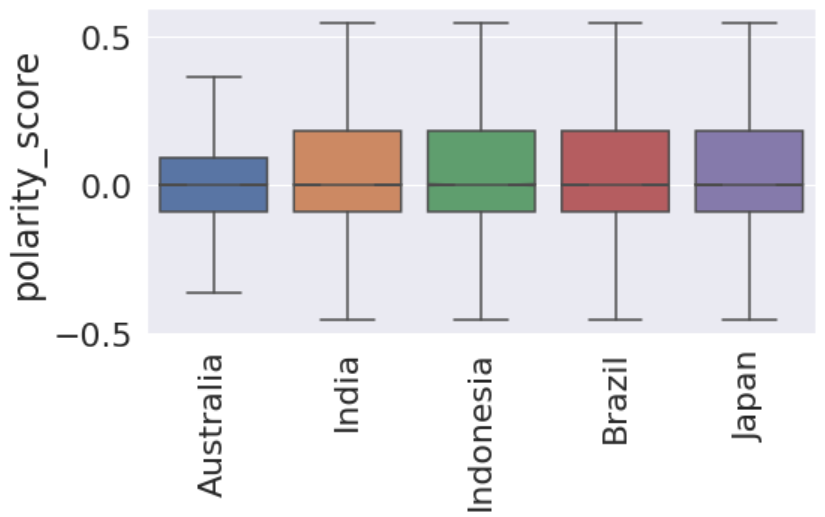}%
\caption{Boxplot analysis of polarity score among different countries}
\label{fig:box} 
\end{figure}

\begin{figure}[htbp!]
\centering 
\   
    \includegraphics[height=4cm]{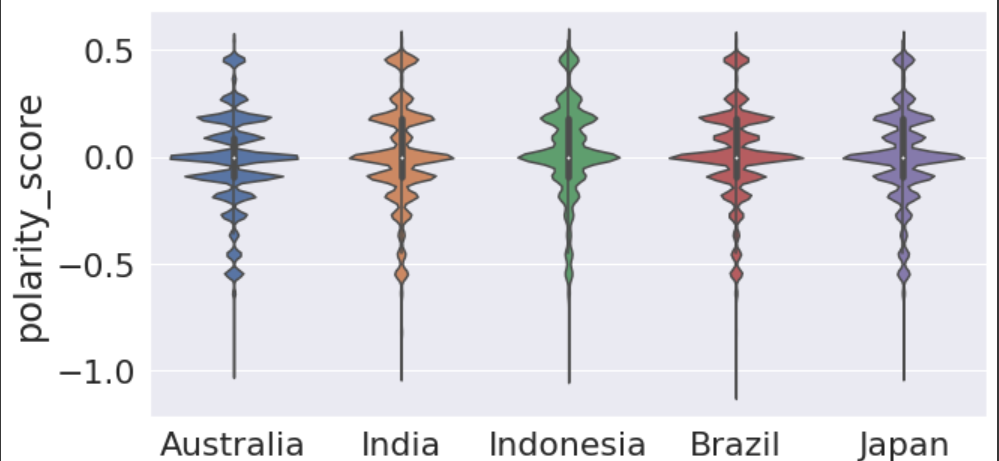}%

\caption{Violin plot of vaccine polarity score for different countries.}
\label{fig:vio} 
\end{figure}

Figure \ref{fig:box}  presents the boxplot that shows that the median of polarity scores is very close to 0 or almost 0. By examining the interquartile ranges for each country, we can discuss any differences in the variability of polarity. A wider interquartile range suggests a greater variation in sentiment among Twitter users in that country, while a narrower range suggests a more consistent sentiment, as in the case of Australia. Figure \ref{fig:vio} shows a violin plot of vaccine polarity scores for different countries that  suggests the polarity density of tweets for different countries.

\begin{figure}[htbp!]
\centering 
\subfigure[ Barplot 1 ]{
    \includegraphics[height=3.5cm]{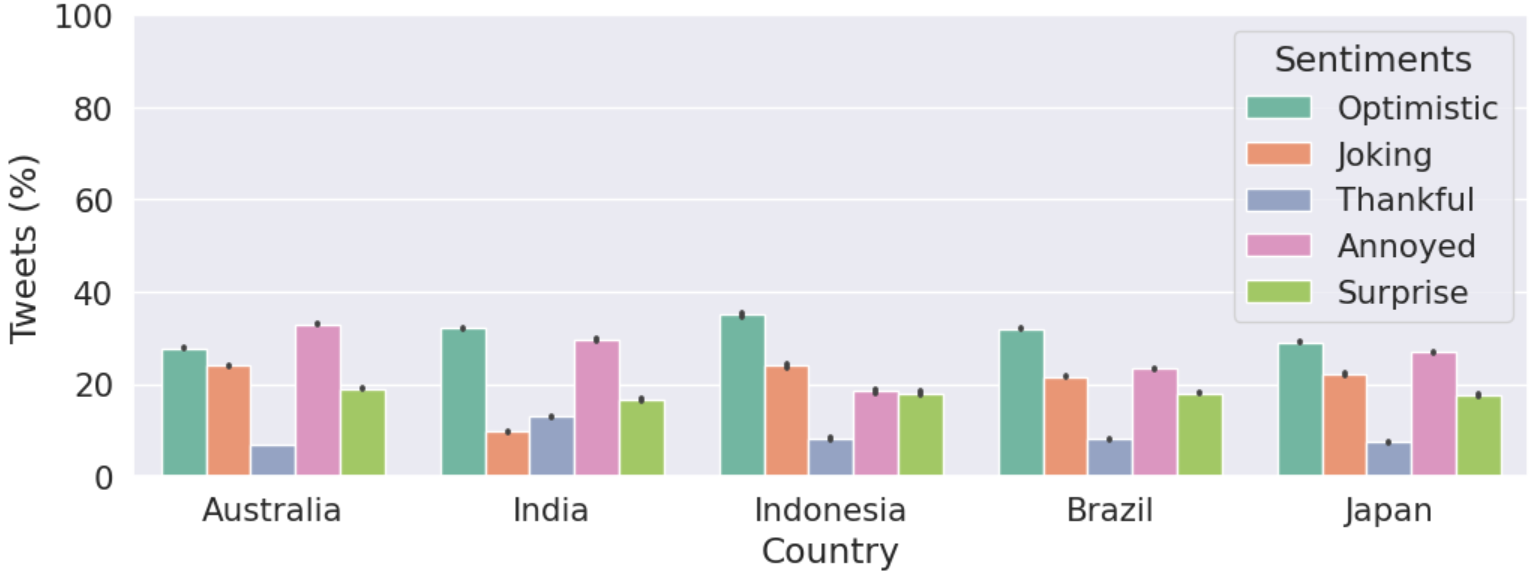}%
}
\subfigure[ Barplot 2 ]{
    \includegraphics[height=3.5cm]{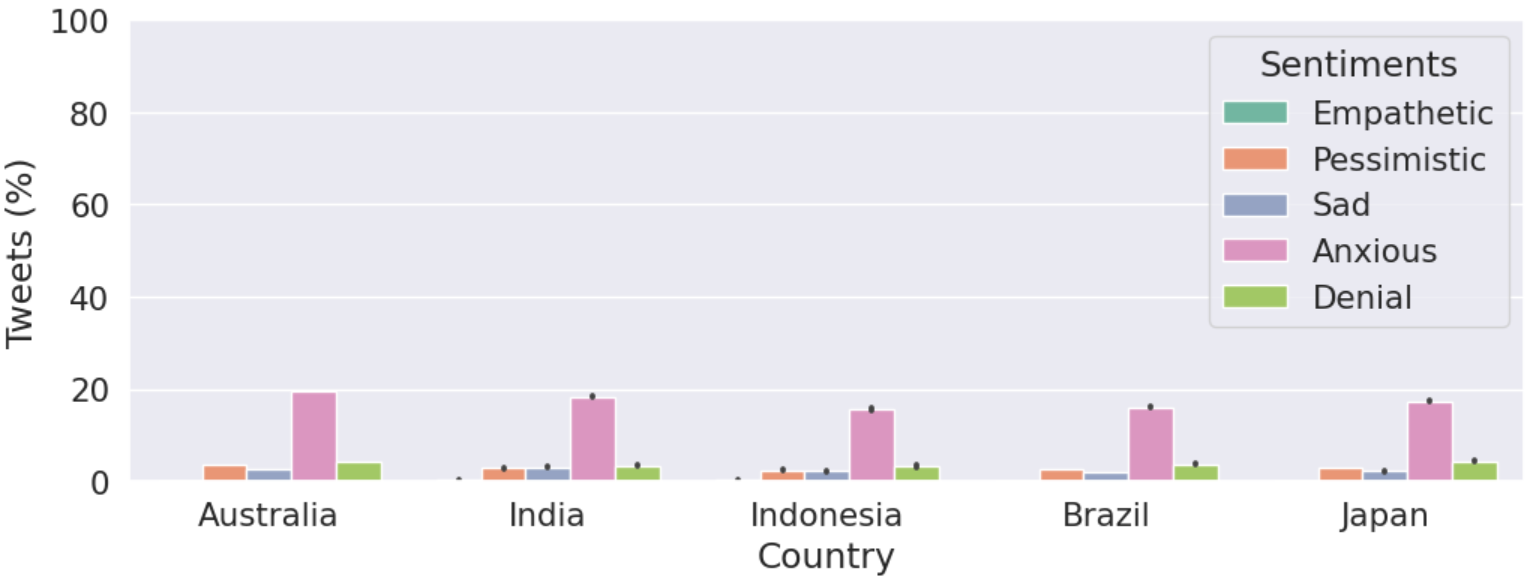}%
}
\caption{ This plot shows the percentage of tweets belonging to sentiments for each country.}
\label{fig:trendx} 
\end{figure}

Figure \ref{fig:trendx} presents the percentage of tweets belonging to sentiments for each country.  We note that optimistic, joking and annoyed are the most expressed sentiments (Panel a) along with anxious (Panel b).

\begin{figure*}[htbp!]
\centering
\subfigure[Mean monthly polarity score obtained from our proposed method (Figure 2)]{
  \includegraphics[height=8cm]{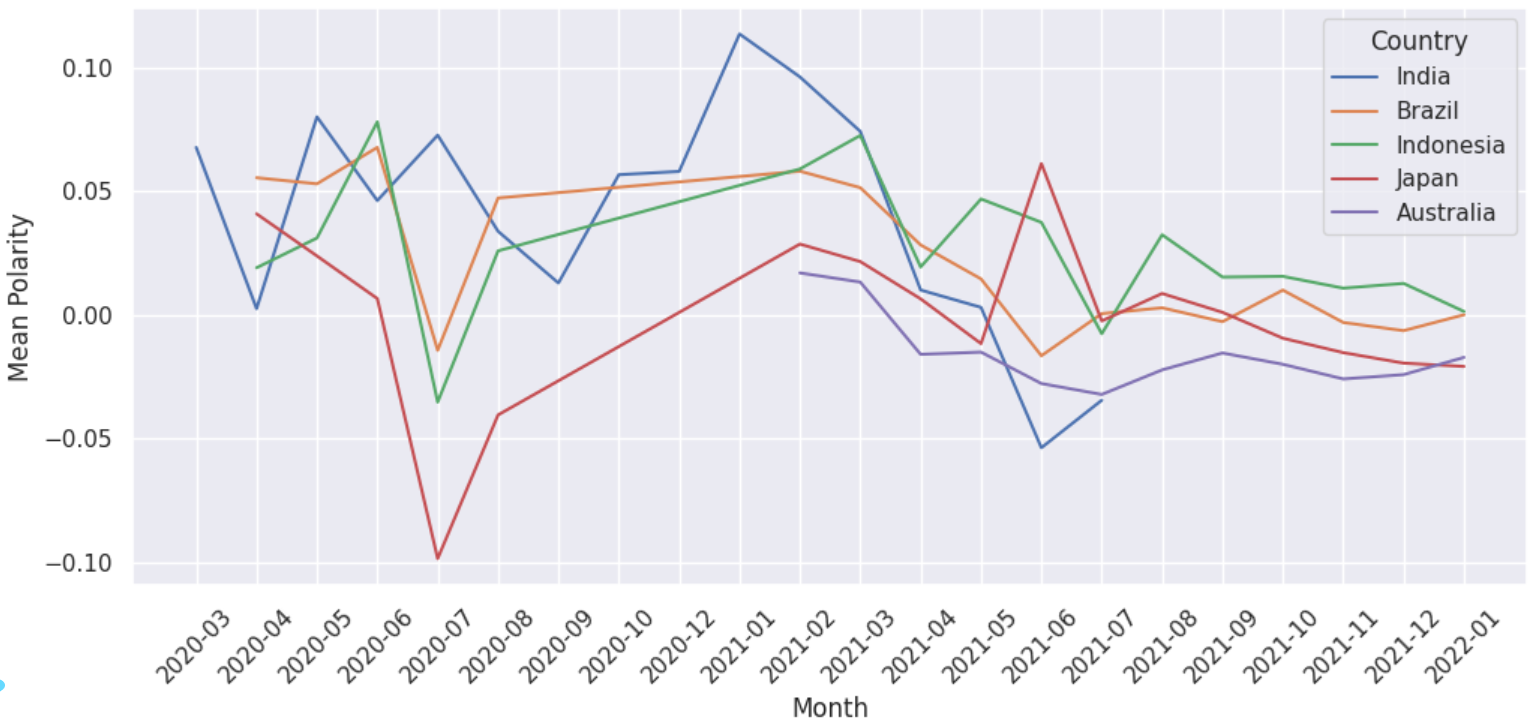}
}
 
\subfigure[Mean monthly  Textblob polarity score]{
  \includegraphics[height=8cm]{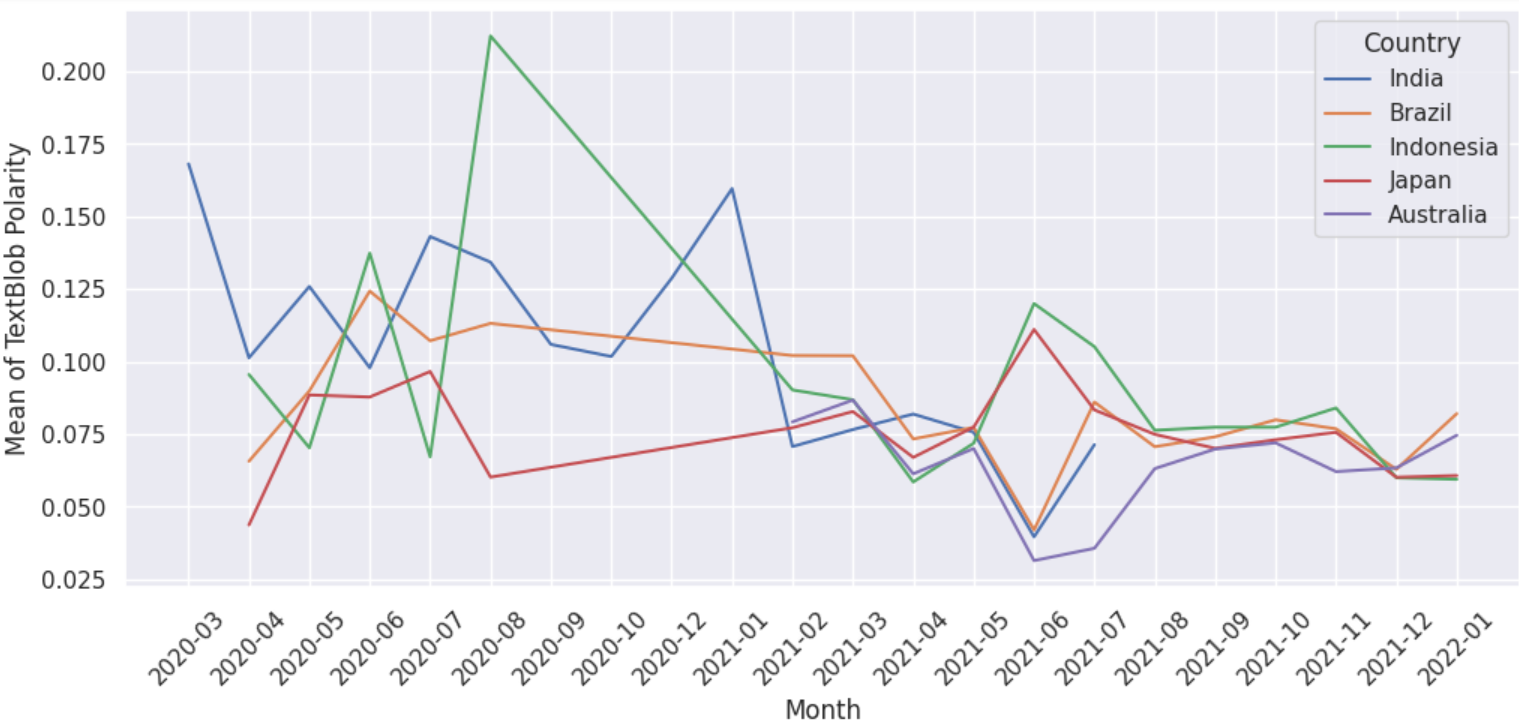}%
}
 
\caption{Monthly  polarity scores of selected countries obtained by TextBlob and our proposed polarity score method shown Figure 2.}
\label{fig:monthlypolarity} 
\end{figure*}

\begin{figure*}[htbp!]
 \centering
    \includegraphics[height=7cm]{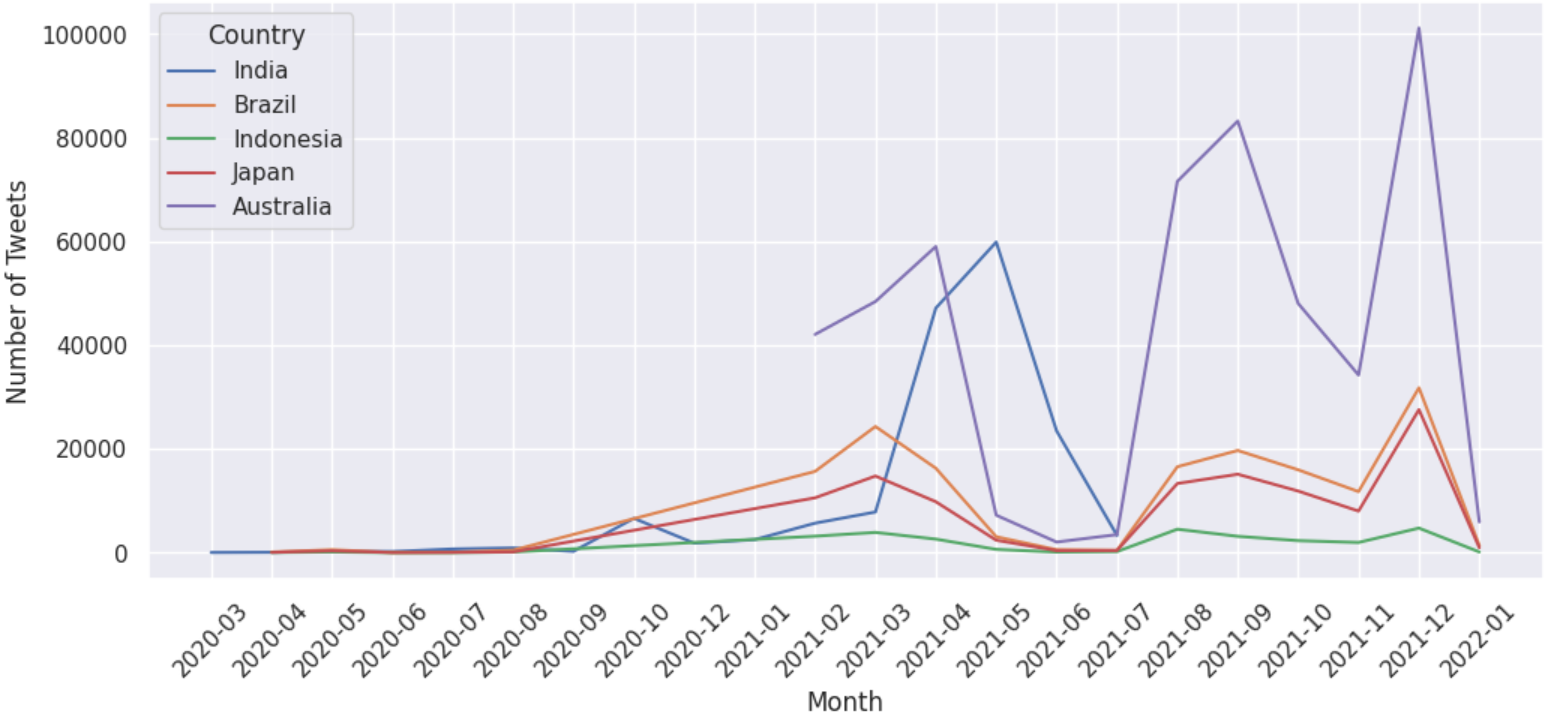}%
\caption{Monthly analysis of the number of tweets related to vaccines for each country}
\label{fig:monthlytweets} 
\end{figure*}

\begin{figure*}[htbp!]
 \centering
    \includegraphics[height=7cm]{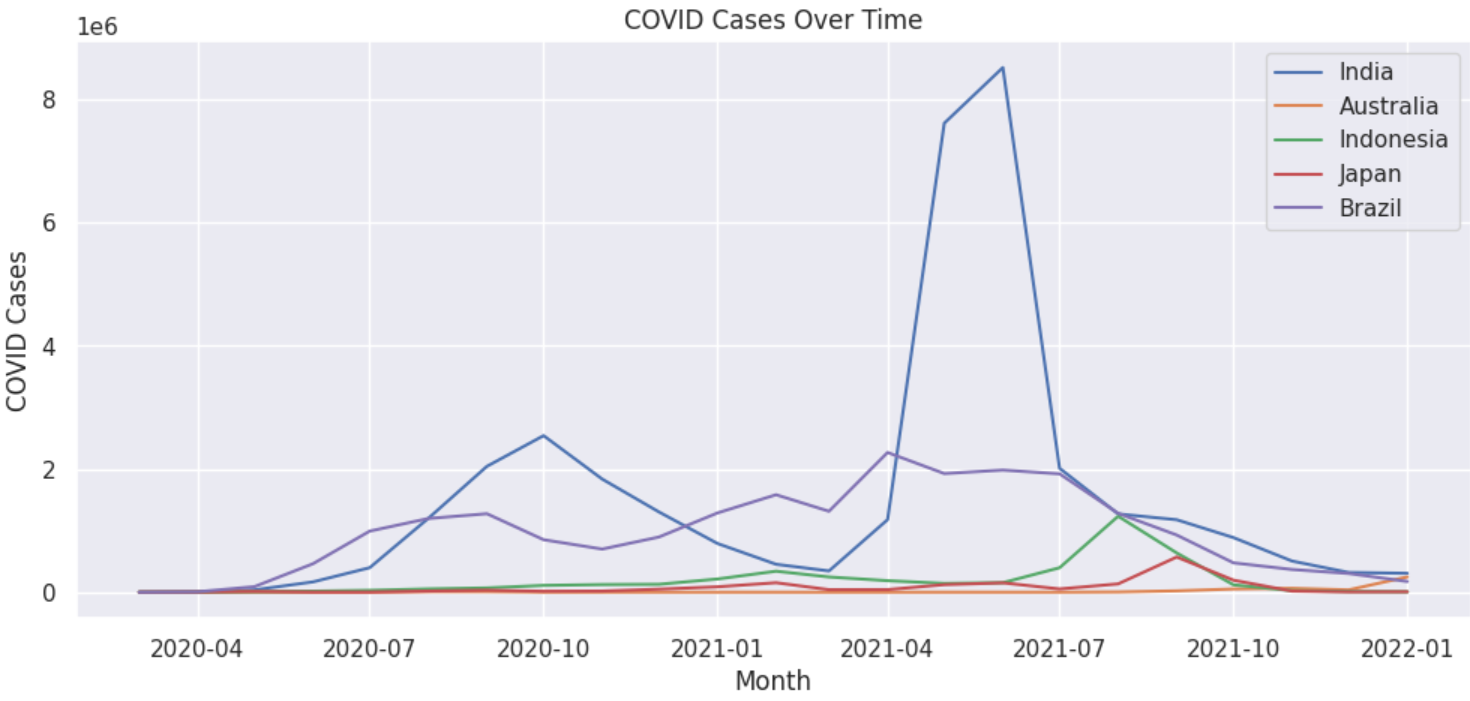}%
\caption{Monthly analysis of the number of COVID-19 cases of each country}
\label{fig:monthycases} 
\end{figure*}

\begin{figure*}[htbp!]
\centering
\subfigure[ February 2021 - July 2021]{
    \includegraphics[height=6cm]{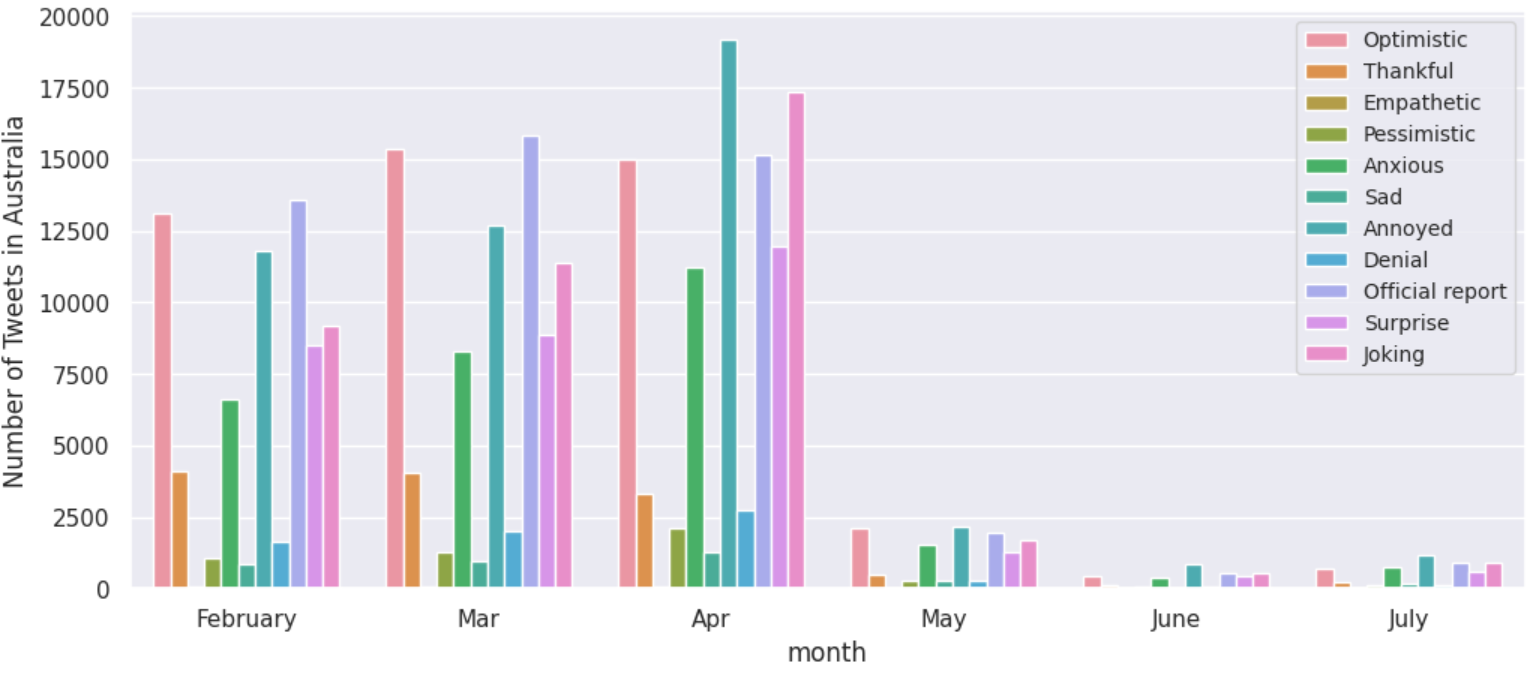}
}
\subfigure[ August 2021 - January 2022]{
    \includegraphics[height=6cm]{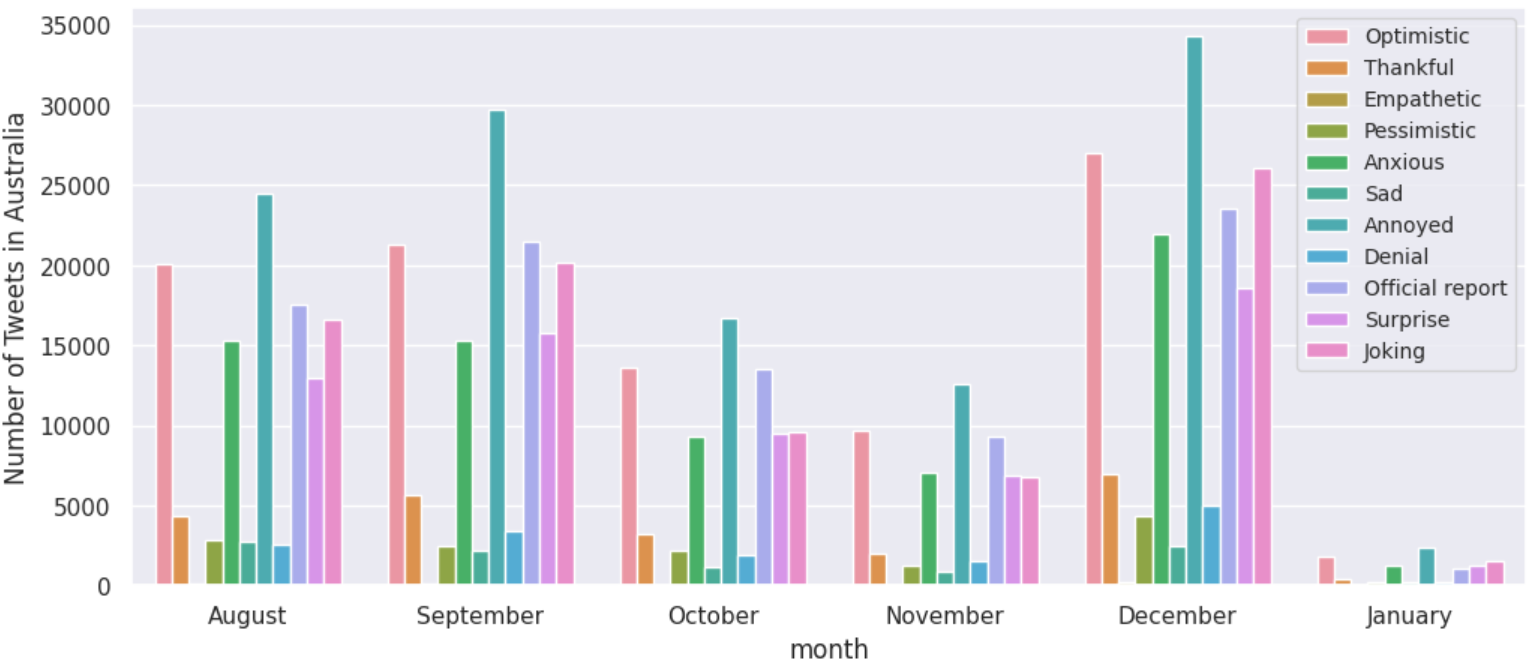}%
}
\caption{ Monthly analysis of tweets for each sentiment for Australia.}
\label{fig:austrend} 
\end{figure*}

We next present the longitudinal analysis of the polarity score, the number of tweets, and the number of cases for the five selected countries.  Figure \ref{fig:monthlypolarity}  presents the monthly  polarity scores  computed by our polarity score (Figure 2)  methodology (Panel a) and TextBlob NLP library 
(Panel b) for each country which overlaps with the number of monthly tweets in  Figure \ref{fig:monthlytweets} and the number of novel monthly cases in Figure \ref{fig:monthycases}.   Overall, we find that both methods (Panel a and Panel b) in Figure \ref{fig:monthlypolarity} have a similar trend from February 2021. In the first half of the pandemic (March 2020 to January 2021), we notice that both methods have drastic changes in the polarity scores. In the case of Japan, Panel a shows that the polarity decreases from April - July 2020, whereas Panel b shows that the polarity increased during the same period. Afterwards, both methods show an increasing trend. In the case of other countries, we see that in majority of the time, there are similar trends given by both methods. 

We notice that India has been one of the worst affected countries by the COVID-19 pandemic (Figure \ref{fig:monthycases}), with two major waves of infections hitting the country. In comparison to the rest of the countries in this study, the population of India is much higher, hence it's natural that there will be more cases during the peak.   The second wave of COVID-19 in India was much more severe than the first wave in terms of the number of cases and deaths. We find that the rest of the countries had a much lower number of cases during their peaks.   Looking at the number of tweets and polarity score, we note that we did not have complete data available for India and hence the study is only till June 2021 (Figure \ref{fig:monthlypolarity} and \ref{fig:monthlytweets}). Moreover, in the case of India, we  notice that both the polarity scores from February 2021 to June 2021 (Figure \ref{fig:monthlypolarity}) decreased drastically (Panel a and Panel b)  with increasing cases from February 2021 (Figure \ref{fig:monthycases}). During this timeline, the number of monthly tweets also increased drastically (Figure   \ref{fig:monthlytweets}). These figures very well capture the situation during the second peak of COVID-19 cases in India, which has a devastating effect in terms of deaths and hospitalisations and lack of resources. We also notice that the polarity score for Japan decreased drastically during the first few months of the pandemic, although there were not many cases. The rise of infections  in other countries could have affected the social media activity of Japan. We also note that in the case of Australia, data was available from February 2021 and we find that the number of tweets drastically increased from July 2021 (Figure   \ref{fig:monthlytweets}) which aligns well with an increase in the number of cases (Figure \ref{fig:monthycases}).

In the second half of  2021, most countries had already started their vaccination programs, and some had even completed the vaccination of their most vulnerable populations. However, the pace of vaccination and the specific groups being targeted for vaccination can vary widely from country to country. So we can observe the effect of vaccination in the mean polarity of the majority of the selected countries (Figure \ref{fig:monthlypolarity}) after April 2021 moves towards the neutral state (mean polarity of 0). This is despite the rise in the number of monthly tweets and cases during this time frame which demonstrates the effect of the vaccination drive.

 We  select  random samples of tweets with their predicted sentiments  and sentiment scores   for Australia and India in Tables 3 and 4, respectively. The rest of the countries are presented in Tables 5 - 7 in the Appendix. We note that despite the fact that certain sentiment labels such as 'anxious", 'annoyed" and 'pessimistic" are conventionally interpreted  as negative emotions (which would imply  anti-vaccine sentiments), we notice the problem with double negative statements which may convey the opposite meaning \cite{baker1970double}.  We notice that the "official report" which is a sentiment label has a polarity score of 0 in the case of Australia (Table 2), and is also labelled as "anxious" in the case of India with 
a negative polarity score (Table 3). The rest of the tweets align with their sentiments captured by the BERT-based model and associated polarity score. Figure \ref{fig:austrend} presents the sentiments expressed in Australia over the course of the pandemic. We notice that "optimistic", "joking", "thankful" and "anxious" are the major sentiments expressed with change in volumes from May to July 2021, which was due to the   roll-out of vaccines, and also a lower number of cases. The rest of the countries are presented in Figures 17 - 20 in the Appendix.

\begin{figure}[htbp!]
\centering
\subfigure[Trigram for negative sentiments for India]{
   \includegraphics[height=4cm]{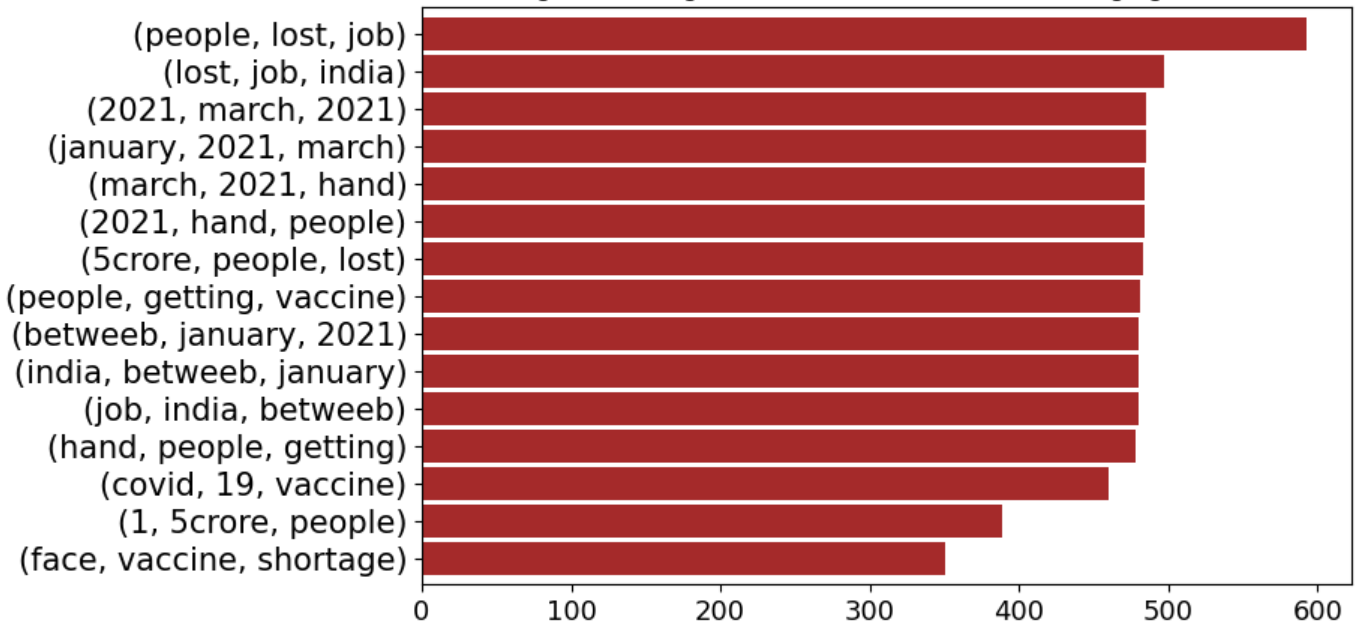}%
}
\subfigure[Trigram for neutral sentiments for India]{
   \includegraphics[height=4cm]{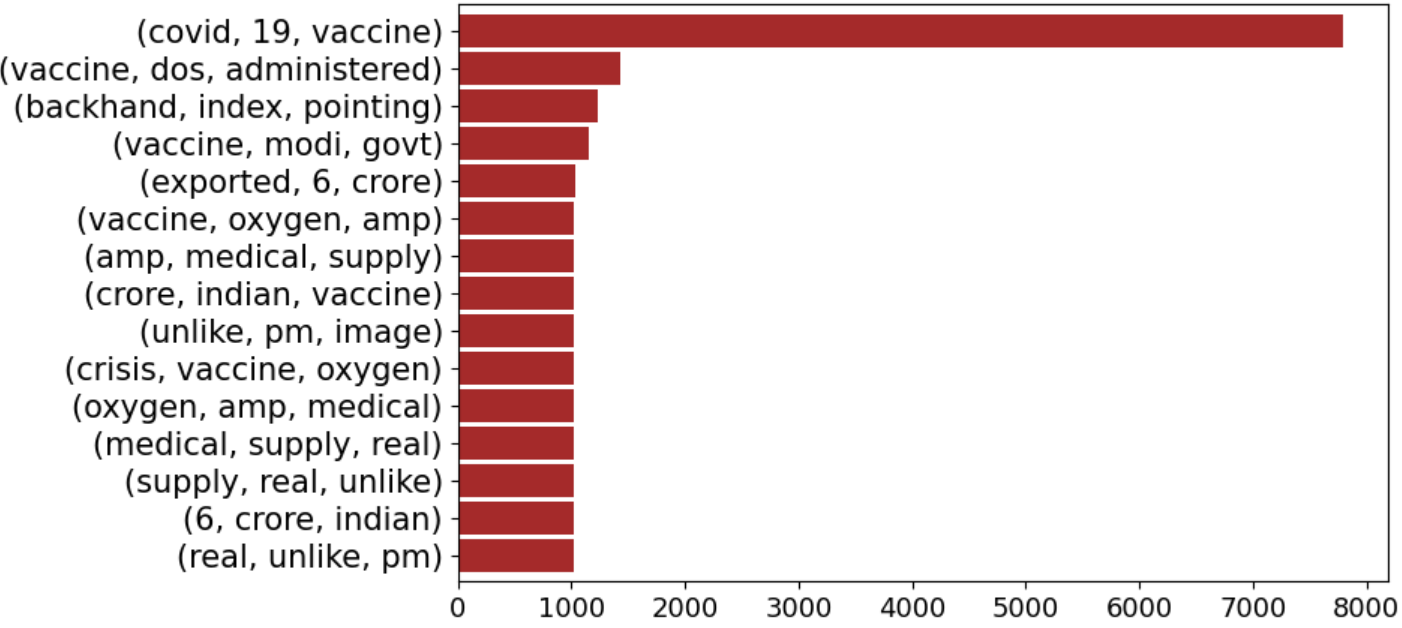}%
}
\subfigure[Trigram for positive sentiments for India]{
   \includegraphics[height=4cm]{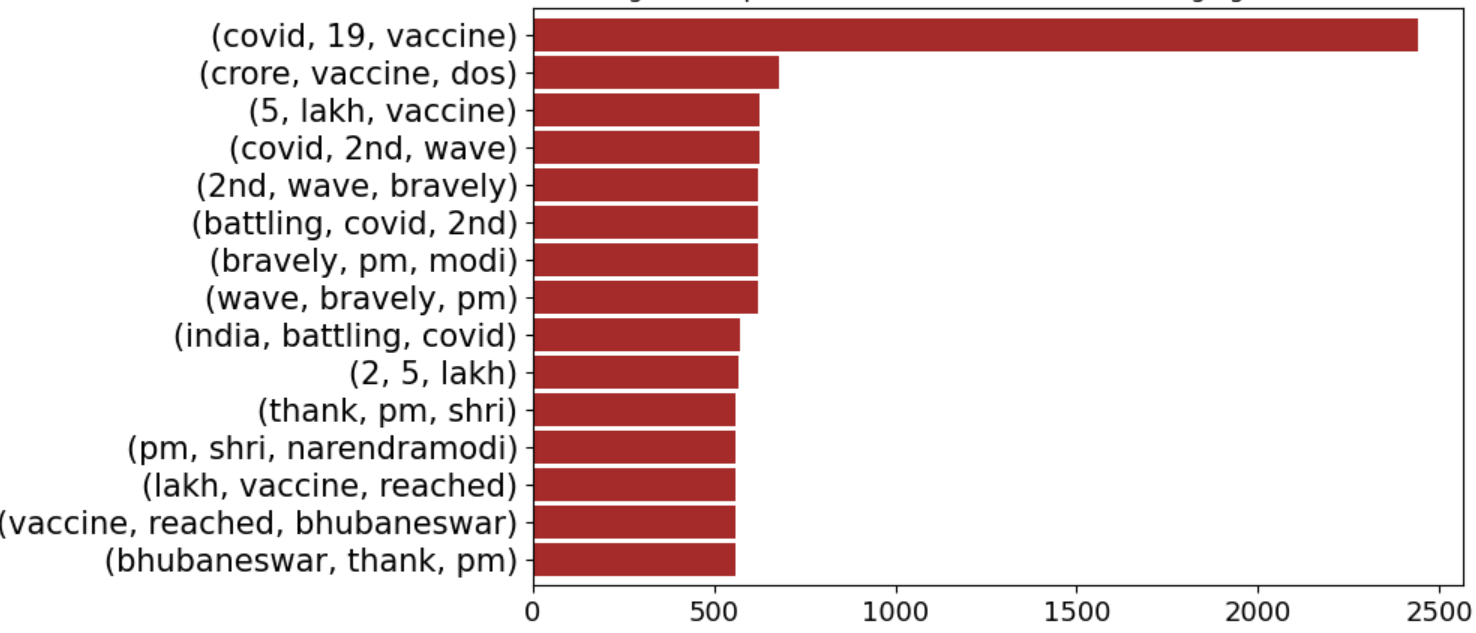}%
}
 \caption{Trigrams associated with India for negative, positive and neutral sentiment scores.}
\label{fig:triIndia}
\end{figure}

\begin{figure}[htbp!]
\centering
\subfigure[Trigram for negative sentiments for Australia]{
   \includegraphics[height=4cm]{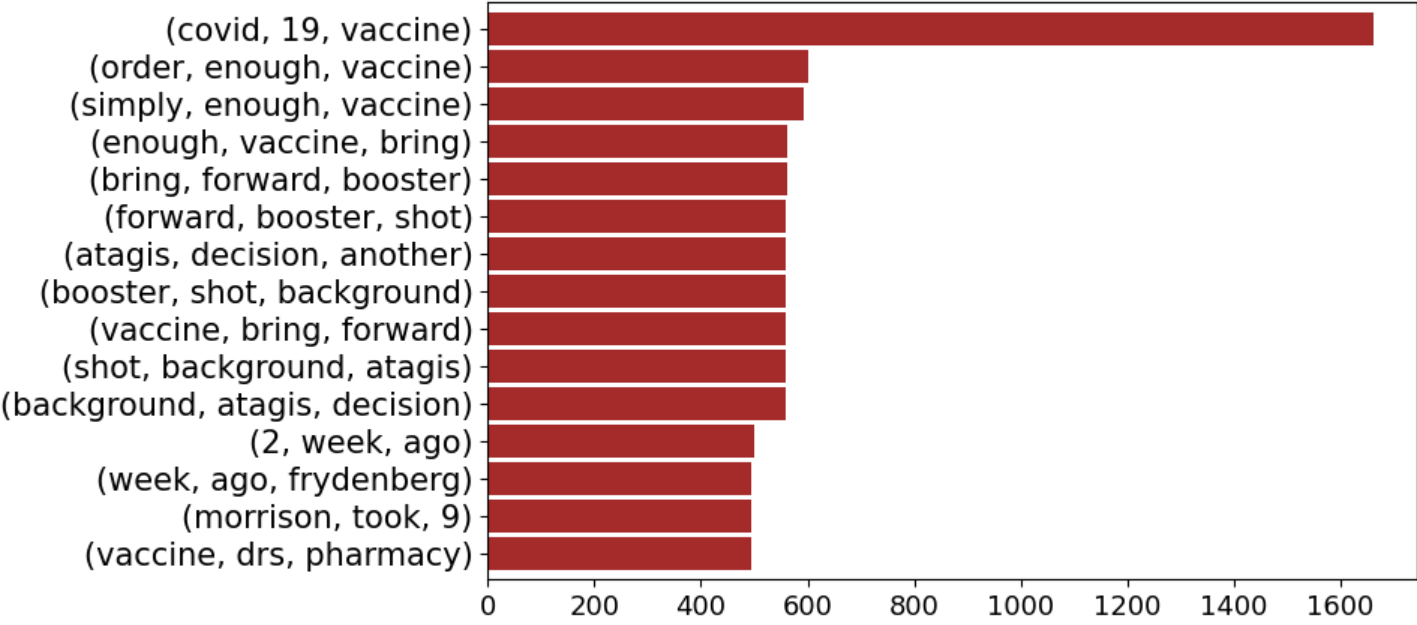}%
}
\subfigure[Trigram for neutral sentiments for Australia]{
   \includegraphics[height=4cm]{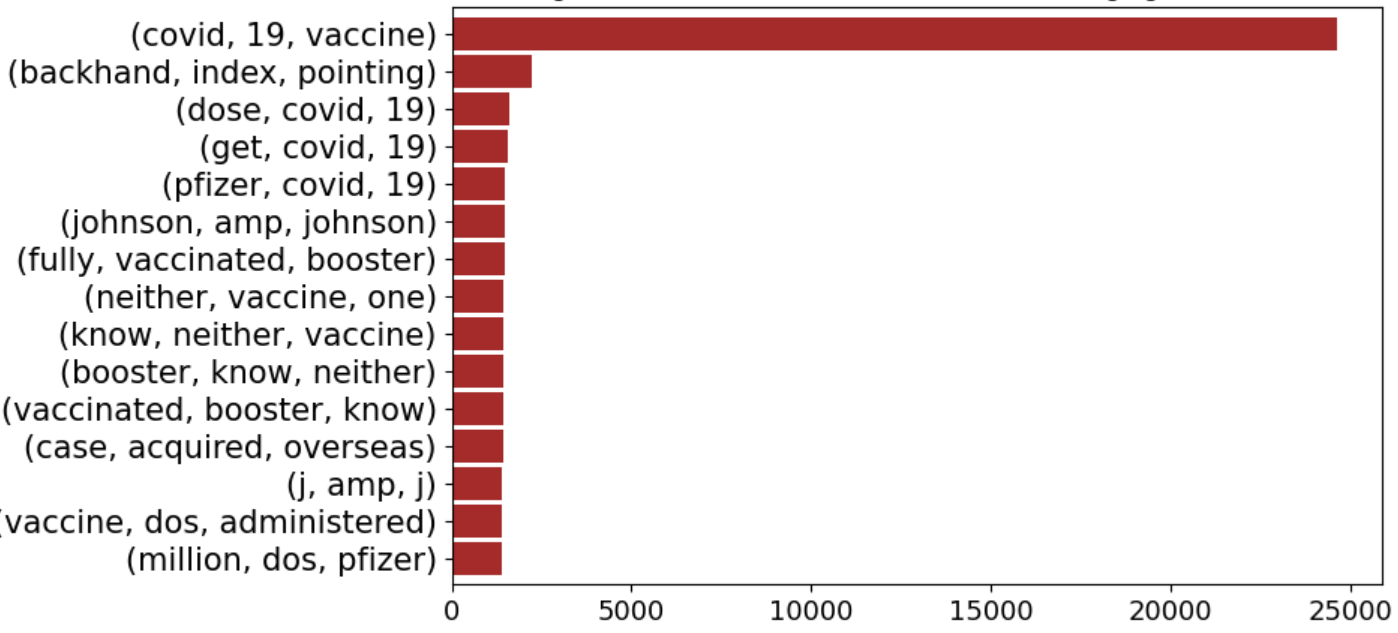}%
}
\subfigure[Trigram for positive sentiments for Australia]{
   \includegraphics[height=4cm]{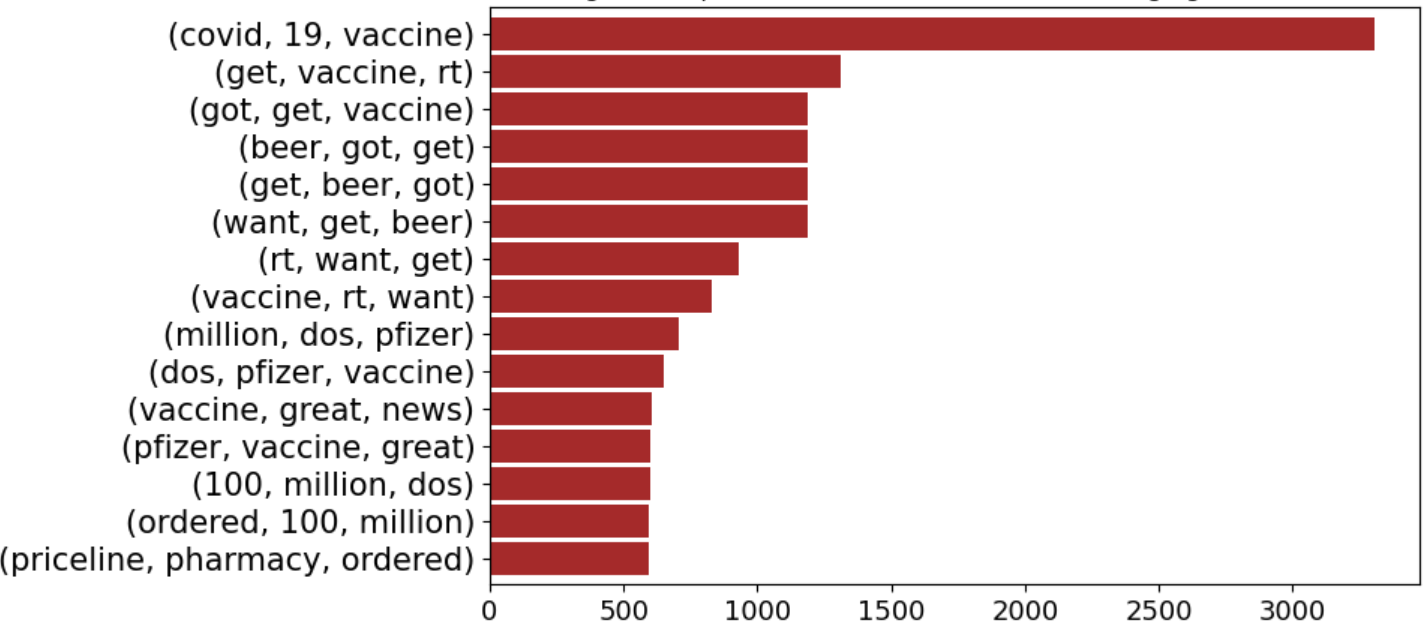}%
}
 \caption{Trigrams associated with Australia for negative, positive and neutral sentiment scores.}
\label{fig:triaus}
\end{figure}

Finally, we present the trigrams for sentiment polarity $p$ based on TextBlob where we group negative sentiments ($p \leq-0.2$), positive sentiments ($p> 0.2$), and neutral sentiments ($p> -0.2$ and $p \leq 0.2$). Figures \ref{fig:triIndia} and \ref{fig:triaus} present the trigrams associated with the three groups of sentiments for the case of India and Australia, respectively. In the case of India (Figure \ref{fig:triIndia}), we find that negative sentiments are associated with "people, lost, job" and topics such as vaccine shortage, while the neural sentiments also relate to vaccines and other issues such as medical supply. Indian Prime Minister Narendra Modi. The positive  sentiments are associated with the COVID-19 vaccine, vaccine dosage, second wave, and Prime Minister Narendra Modi. There are certain topics (trigrams) that overlap negative and neutral sentiments and overall all the sentiments. Vaccine overlaps all three sentiment groups, which indicates that based on the context, groups of people expressed it as negative, positive and neutral. Figure \ref{fig:triaus} presents the trigrams for the case of Australia where we find vaccine as the most expressed trigram across the three sentiment groups. This is similar to India and we also find that Promie Minister Scott Morrisson amongst the  trigrams of the negative sentiment group which Indicates majority of tweets expressed negative sentiments about how the government managed the pandemic. The  names of vaccine manufacturers  such as Pfizer and Johnson and Johnson, are among the neutral sentiments. We find trigrams associated with vaccines and their manufacturers, order numbers etc, in the group of positive trigrams.

 \begin{table*}[htbp!]
    \small
     \centering
    \begin{tabular}{p{4cm}p{8cm}p{2cm}p{1cm}}
    \hline
    \textbf{Sentiments} &  \textbf{Sample tweet} &  \textbf{Country} & \textbf{Score}\\ 
    \hline
     Annoyed &  \textit{" Oh wow. So anti-vaxxers in Melbourne are protesting the salvation army because they want to be able to enter op shops."} & Australia & -0.09 \\%
    \hline
    Annoyed, Anxious &  \textit{"I think boosters at 3 months might be better for now with future testing to determine efficacy Much talk about the vaccines requiring 3rd shot to supply full immunity. This is not unusual amongst vaccines however it is being used by anti-vaxxers to undermine the vaccine rollout."} & Australia  & -0.27 \\%
    \hline
    Joking, Surprise & \textit{"Which one are you ? ... covidiots and antivaxxers."} & Australia & 0.09 \\
    \hline
    Optimistic, Thankful & \textit{"Mr ... I like that story of the GP phoning up any of patients who hadn't vaccinated saying "You never doubted any of the other vaccines or medicines I've prescribed for you." He has 70 percent success rate in getting antivaxxers vaccinated. knighthood"} & Australia & 0.45 \\
    \hline
    Official report & \textit{"RT The D I C: Pandemic laws contain most rigorous safeguards in nation, say experts."} & Australia & 0.00\\
    \hline
    \end{tabular}
    \smallskip
    \caption{Sentiment prediction outcomes for randomly selected tweet samples for Australia and their sentiment scores. }
    \label{tab:antivaxx-score}
\end{table*}

\begin{table*}[htbp!]
    \small
     \centering
    \begin{tabular}{p{4cm}p{8cm}p{2cm}p{1cm}}
    \hline
    \textbf{Sentiments} &  \textbf{Sample tweet} &  \textbf{Country} & \textbf{Score}\\ 
    \hline
     Optimistic &  \textit{" Good news!!! Researchers are saying that the Corona virus is not mutating...Which means a vaccine development agai…."} & India & 0.18 \\%
    \hline
    Joking &  \textit{" I'd say it was karma catching up with an anti vaxxer corona spreader, but karma would be remiss in…"} & India  & 0.09 \\%
    \hline
    Anxious, Official & \textit{"117 million + Children risk missing out on  measles vaccines, as COVID19 surges: MeaslesRubella"} & India & -0.18 \\
    \hline
    Annoyed, Denial & \textit{"Is china cheated whole world, how 74k patient recovered from out of 81k, without the vaccine, without full Lockdown…"} & India & -0.54 \\
    \hline
    Empathetic & \textit{"There is. Make vaccine for corona virus...! Let’s pray for it..."} & India & 0.00\\
    \hline
    \end{tabular}
    \smallskip
    \caption{Sentiment prediction outcomes for randomly selected tweet samples for India and their sentiment scores. }
    \label{tab:antivaxx-score}
\end{table*}

 \section{Discussion}


    Our study about the COVID-19 vaccine sentiment analysis study has provided insights into the dynamic moment of sentiments towards vaccines during a global pandemic. Our results show that the volume and sentiment polarity of tweets is closely related to the phase of the pandemic, i.e there the selected countries experienced drastic changes in the polarity sore at the beginning of the pandemic, which stabilised at the second half of the pandemic (Figure \ref{fig:monthlypolarity}.  It is evident that local and global peaks in COVID-19 vaccine-related tweets were identified from May to June 2020 and from March to May 2021, which correlates with the roll-out of clinical trials \cite{oldenburg2021diagnosis} and major concerns for the safety and efficacy of the Astra-Zeneca vaccine \cite{knoll2021oxford}. There was massive fear in social media due to  blood clots being developed by patients around the world from the Astra-Zeneca vaccine which was then suspended by European countries \cite{wise2021covid}.

Vaccine hesitancy has been a major global public health threat, especially during the COVID-19 \cite{dror2020vaccine,dube2021vaccine}.
We find more tweets being identified with negative polarity scores when vaccine development progress has been announced and when transmission has been reasonably contained in the case of Australia (Figure \ref{fig:austrend}).  Figure 9 shows that all the selected countries had "Optimistic", "joking", annoyed", "surprised" and "anxious", as the major sentiments expressed. This trend seems to be present throughout the pandemic looking at the case of Australia (Figure \ref{fig:austrend}).

Furthermore, in Figure 4, we notice that 86.7\% of input tweets were assigned with only one or two sentiment labels  which are consistent with our prior studies about COVID-19 sentiment analysis for India  \cite{chandra2021covidsentiment}. Three or more sentiments are rarely expressed in everyday speech and given that tweets have an upper limit in the number of words, the predictions by the model make sense which can also be verified by the tweet samples in Tables 2 and 3. These tables provide useful insights for building a better understanding of the vaccine sentiments and the driving forces behind those sentiments. Further work in the area of psychology can be done with post analysis of tweets and the sentiments expressed. Nevertheless, we note that Twitter's upper limit for tweet length of fewer than 280 characters may hinder the ability of the model to discover complex interrelationships among multiple sentiment labels.

We revisit the sample tweets in Table 2 for Australia, we draw attention to the second tweet entry \textit{"I think boosters at 3 months might be better for now with future testing to determine efficacy Much talk about the vaccines requiring 3rd shot to supply full immunity. This is not unusual amongst vaccines however it is being used by anti-vaxxers to undermine the vaccine rollout."} which was labelled as "annoyed" and "anxious" with a polarity score of  -0.27. Double negative statements \cite{baker1970double}, irony, sarcasm  and many more language usages still pose challenges for NLP  \cite{potamias2020transformer,
verma2021techniques}.At times, it is difficult to get the context of the actual tweet, and simply concluding from the sentiment labels and scores can result in misleading interpretations of anti-vaccine sentiments.
In Table 3, \textit{"117 million+ Children risk missing out on measles vaccines,
as COVID-19 surges: MeaslesRubella”} was labelled as "official" and "anxious" with a negative polarity. The wording of official tweets needs to take the anti-vaxxer viewpoints into account, during the pandemic, official statements were misinterpreted and twisted to suit anti-vaxxer narratives. This sends out an important message to  disease control bodies that they should increase transparency and exercise caution and use disseminate information in a timely manner  to minimise the spread of misinformation about vaccines. This needs to be done in  an effort to restore faith in scientific evidence and reduce ungrounded anti-vaccine sentiments on social media. The policy about masks and distancing \cite{dzisi2020adherence,} was also implemented in an ad-hoc manner which created further misinformation, along with the implementation of vaccination, risks and efficacy \cite{kaplan2021influence}.

There are certain limitations to our framework and study. Firstly, the BERT-based model was trained using the SenWave dataset which contained  10,000 sentiment labels manually labelled by experts. This is perceived to be a subjective activity due to individual differences in how each sentiment is perceived. Secondly, the current study is only concerned with user-generated data retrieved from one social media platform, Twitter; where the user demographic is significantly different from the demographic of users on other social media platforms such as Linkedin and Facebook \cite{singh2019analysis}. The complexity and degree of formality of textual inputs can also vary to a significant degree and with COVID-19, Twitter suspended many accounts along with Facebook to limit anti-vaxxer activities \cite{ferrara2020misinformation}.

\section{Conclusions and Future Work}

In this paper, we used novel  language models for vaccine-related sentiment analysis  during the COVID-19 pandemic. We covered major phases of the pandemic in our longitudinal analysis with and without vaccination and different levels of lockdowns and economic activity that shifted the topics of discussions on social media. Our analysis shows that there is a link between the number of tweets,  the number of cases, and the change in sentiment polarity scores during major waves of COVID-19 cases, especially in the case of India and Australia. We also found that the first half of the pandemic (March 2020 - June 2021) had drastic changes in the sentiment polarity scores which later stabilised, although the number of cases and tweets was high afterwards. This implies that the vaccine rollout had an impact on the nature of discussions on social media - with a more positive outlook toward the pandemic. 

Further work can be carried out to implement   sentiment analysis for vaccinations taking into account  cultural and religious beliefs,  political, and economic landscapes. Furthermore, we can extend this framework and include topic modelling to have a better understanding of the range of topics that emerged with anti-vaccine and prop-vaccine tweets.

\section*{Code and Data}

GitHub repository for this project: \footnote {https://github.com/sydney-machine-learning/COVID19-antivaccine-sentimentanalysis}. 


\section*{Acknowledgements}

We thank Yashwant Singh Kaurav from the Indian Institute of Technology Delhi, India.

\bibliographystyle{IEEEtran}
\bibliography{usyd,Chandra-Rohitash,Bays,2018,aicrg,rr,sample,sample_,2020June,covid,glove,extra_covid,language}








\section*{Appendix}

\begin{table*}[htbp!]
    \small
     \centering
    \begin{tabular}{p{4cm}p{8cm}p{2cm}p{1cm}}
    \hline
    \textbf{Sentiments} &  \textbf{Sample tweet} &  \textbf{Country} & \textbf{Score}\\ 
    \hline
     Optimistic &  \textit{"RT Thankfully, despite the antivaxxers, Reefton is at 85.5 percent first dose, with needing only 40 more people vaccinated to get…"} & Japan & 0.18 \\%
    \hline
    Joking &  \textit{"RT wagatwe: Imagine letting your kids die to stay married to an antivaxxer"} & Japan  & 0.09 \\%
    \hline
    Annoyed, Official & \textit{" UPDATE: William Hartmann, the infamous Wayne Co. canvasser who caused a stir in Michigan by flip flopping on his vote…"} & Japan & -0.09 \\
    \hline
    Joking, Sad & \textit{"So excited for you!!!! Sadly the combo of Omicron +  bad news about the lt 5s pfizer vaccine today is  for parents of babies who can't yet wear masks or get jabbed"} & Japan & -0.18 \\
    \hline
    Pessimistic, Surprise & \textit{" The next 5-10 years online are going to be a period in which the surviving children of covid-antivaxxers get accounts for t…"} & Japan & -0.36\\
    \hline
    \end{tabular}
    \smallskip
    \caption{Sentiment prediction outcomes for randomly selected tweet samples for Japan and their sentiment scores. }
    \label{tab:antivaxx-score}
\end{table*}

\begin{table*}[htbp!]
    \small
     \centering
    \begin{tabular}{p{4cm}p{8cm}p{2cm}p{1cm}}
    \hline
    \textbf{Sentiments} &  \textbf{Sample tweet} &  \textbf{Country} & \textbf{Score}\\ 
    \hline
     Optimistic, Thankful &  \textit{" Shout out amp; respect to all hospital amp; health workers who must treat every incoming sick antivaxxer with civility.  You’re su…"} & Brazil & 0.45 \\%
    \hline
    Joking, Pessimistic, Anxious &  \textit{"Everyone is talking about covid, but I'm honestly terrified that antivaxxers might move on from just covid scepticism and graduate to full vaccine antivaxxers. Measles and polio will return because of the incredible stupidity of the human race."} & Brazil  & -0.45 \\%
    \hline
    Surprise, Official & \textit{"UKHSA update "Myocarditis and pericarditis after COVID-19 vaccination" Original"myocarditis ... has been described in a high percentage of children admitted to hospital"Is the update clearer? Will antivaxxers understand it."} & Brazil & 0.00 \\
    \hline
    Denial, Anxious & \textit{"I wonder if this man really  died from  ccp 19 even after taken his astrazeneca vaccine or maybe he was killed to keep the cover over that deadly vaccine and how unsafe it can be for people?"} & Brazil & -0.63 \\
    \hline
    Official, Sad & \textit{" UPDATE: Doug Ericksen has been found. He died from COVID."} & Brazil & -0.27\\
    \hline
    \end{tabular}
    \smallskip
    \caption{Sentiment prediction outcomes for randomly selected tweet samples for Brazil and their sentiment scores. }
    \label{tab:antivaxx-score}
\end{table*}

\begin{table*}[htbp!]
    \small
     \centering
    \begin{tabular}{p{4cm}p{8cm}p{2cm}p{1cm}}
    \hline
    \textbf{Sentiments} &  \textbf{Sample tweet} &  \textbf{Country} & \textbf{Score}\\ 
    \hline
     Optimistic, Thankful &  \textit{" I don't get tankie-type lefties who defend anti-vaxxers and bemoan vaccine mandates - the dramatic success of the early "} & Indonesia & 0.45 \\%
    \hline
    Joking, Sad, Annoyed &  \textit{" Heartbreaking last text that anti-vaxxer bodybuilder sent before Covid killed him"} & Indonesia  & -0.27 \\%
    \hline
    Surprise, Joking & \textit{" Been chatting to country people who have not had vaccine jab. They are not antivaxxers. As they have had no Covid in…"} & Indonesia & 0.09 \\
    \hline
    Empathetic, Optimistic & \textit{"RT its me your mom: antivaxxer: God, why won't you heal me of Covid god: I gave you like 4 vaccines to choose from"} & Indonesia & 0.18 \\
    \hline
    Official, Sad & \textit{" Antivaxxer kickboxing champion Frederic Sinistra, known as "The Undertaker," died in his home in Ciney, Belgium, from…"} & Indonesia & -0.27\\
    \hline
    \end{tabular}
    \smallskip
    \caption{Sentiment prediction outcomes for randomly selected tweet samples for Indonesia and their sentiment scores. }
    \label{tab:antivaxx-score}
\end{table*}

\begin{figure*}[htbp!]
\centering 
\subfigure[ March 2020 - November 2020]{
    \includegraphics[height=7cm]{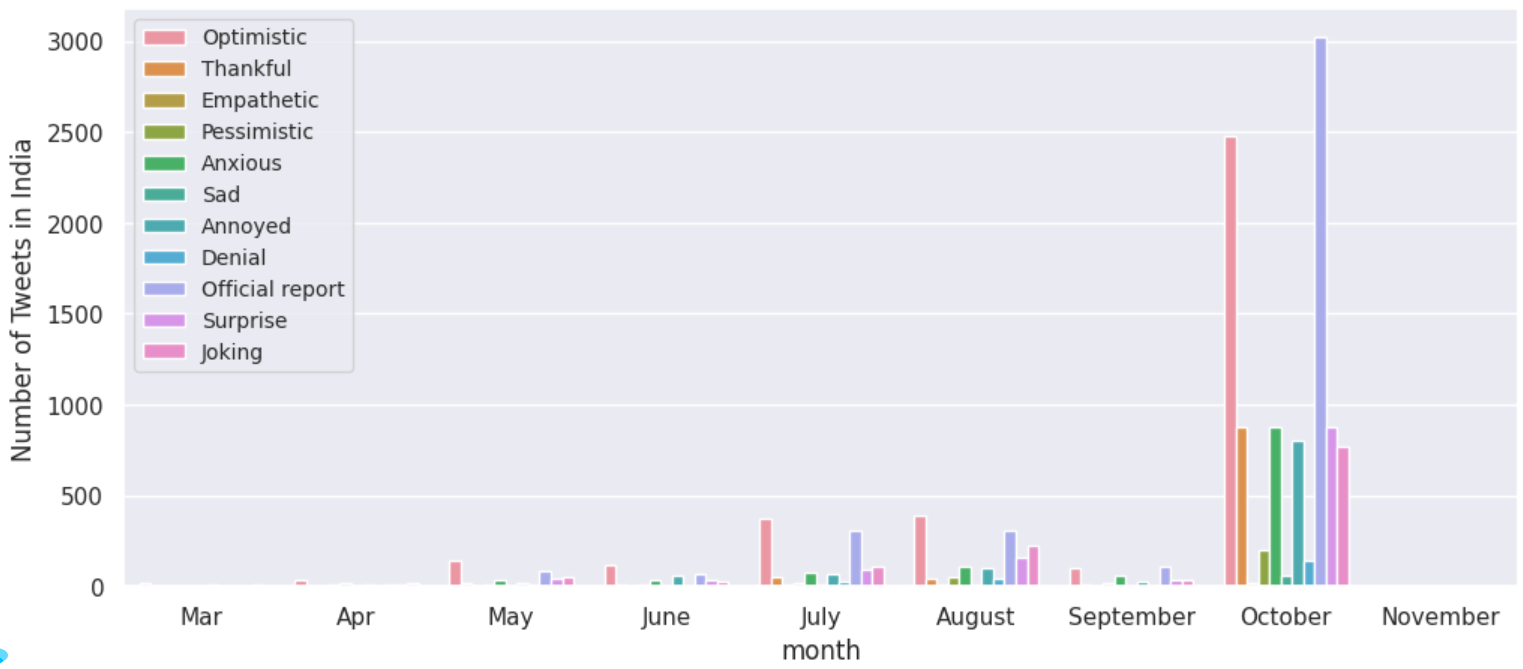}%
}
\subfigure[ December 2020 - July 2021]{
    \includegraphics[height=7cm]{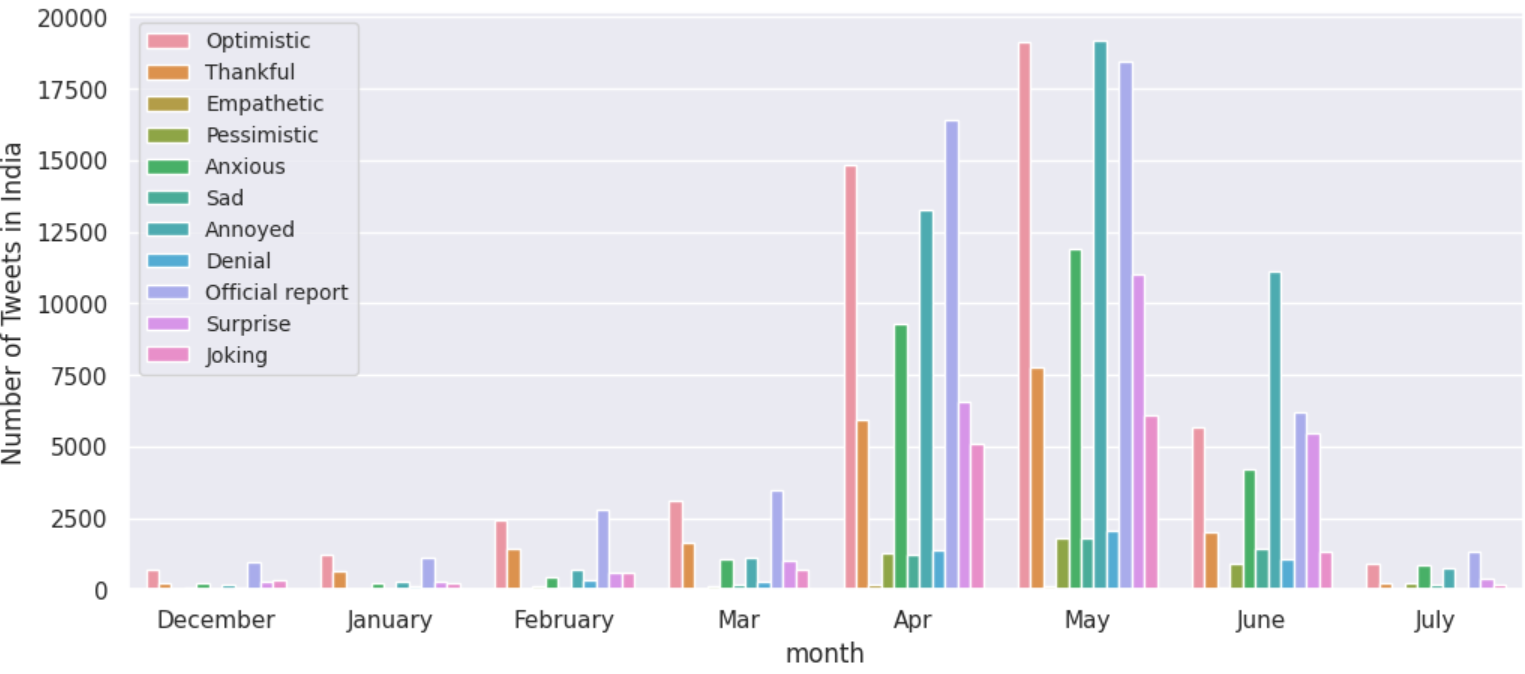}%
}
\caption{ Monthly analysis of the number of tweets for each sentiments for India}
\label{fig:trend} 
\end{figure*}

\begin{figure*}[htbp!]
\centering 
\subfigure[ March 2020 - November 2020]{
    \includegraphics[height=7cm]{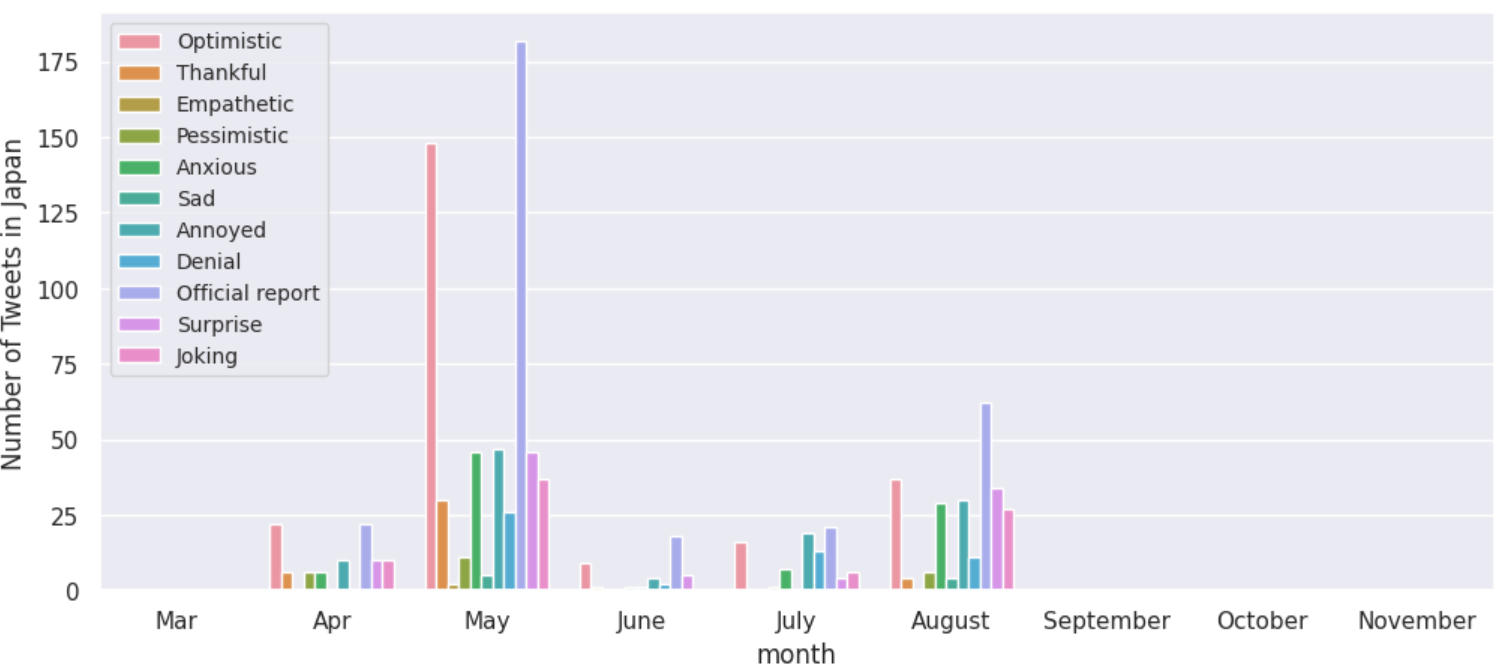}%
}
\subfigure[ December 2020 - July 2021]{
    \includegraphics[height=7cm]{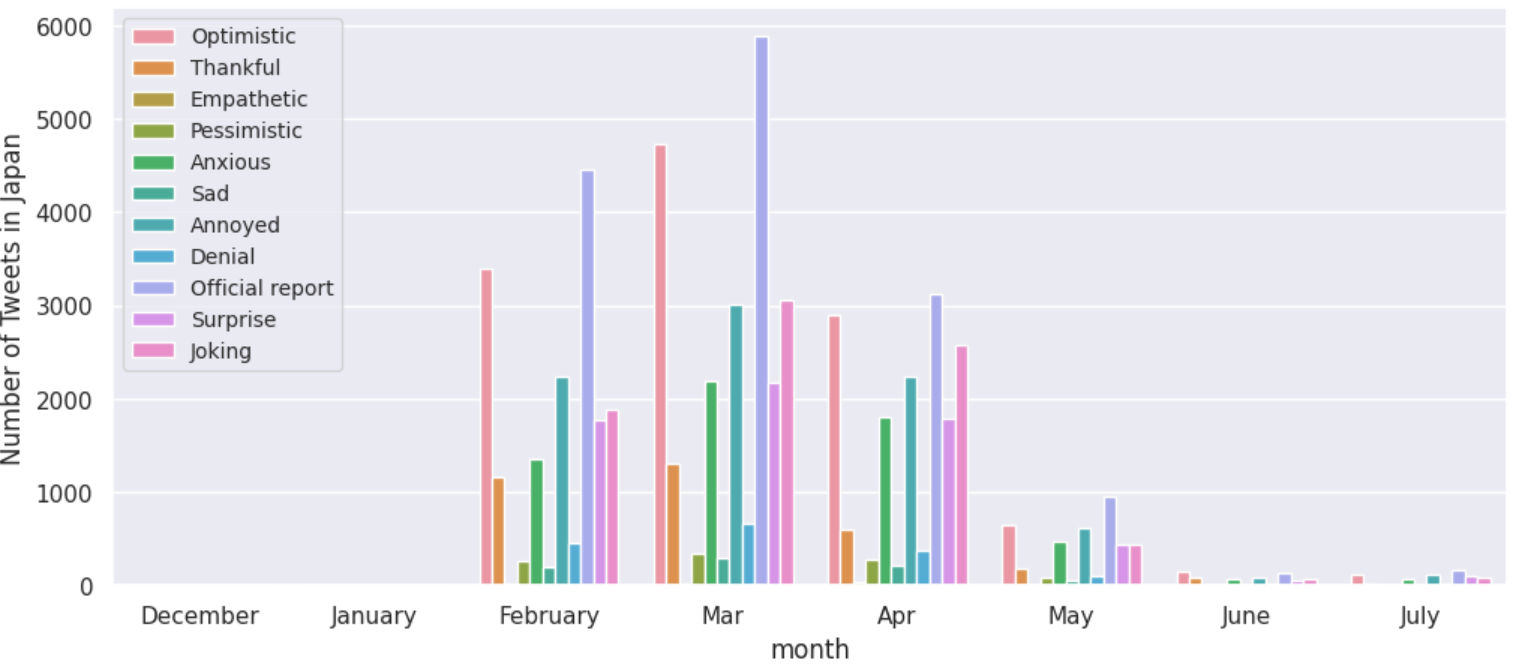}%
}
\subfigure[ August 2021 - January 2022]{
    \includegraphics[height=7cm]{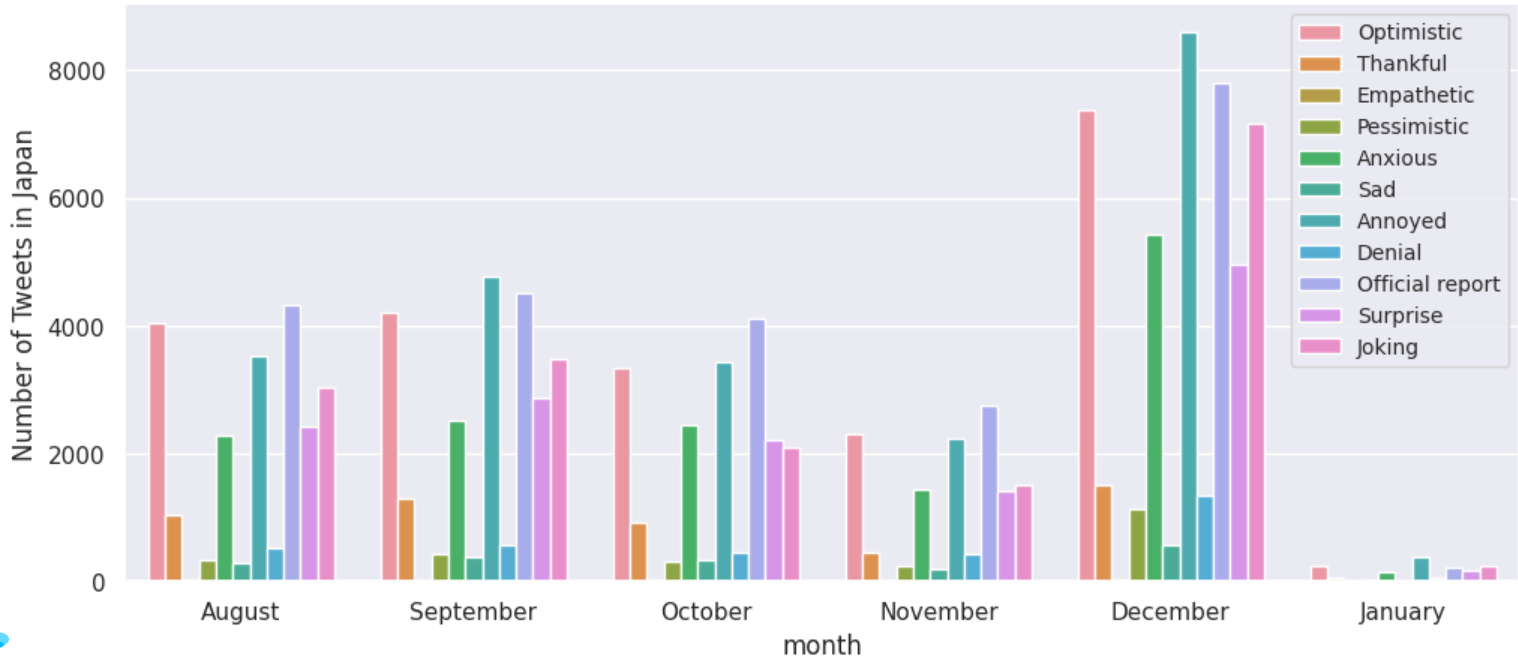}%
}
\caption{ Monthly analysis of the number of tweets for each sentiment for Japan}
\label{fig:trend} 
\end{figure*}

\begin{figure*}[htbp!]
\centering 
\subfigure[ March 2020 - November 2020]{
    \includegraphics[height=7cm]{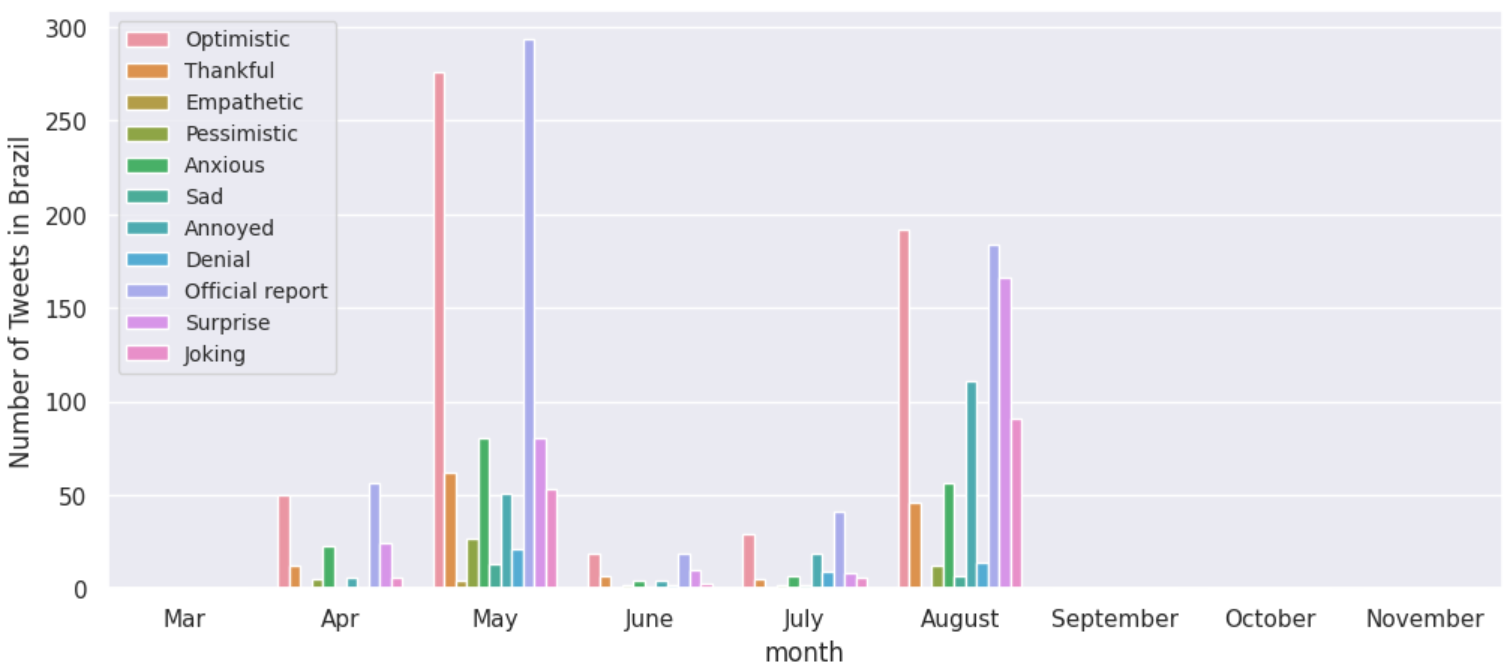}%
}
\subfigure[ December 2020 - July 2021]{
    \includegraphics[height=7cm]{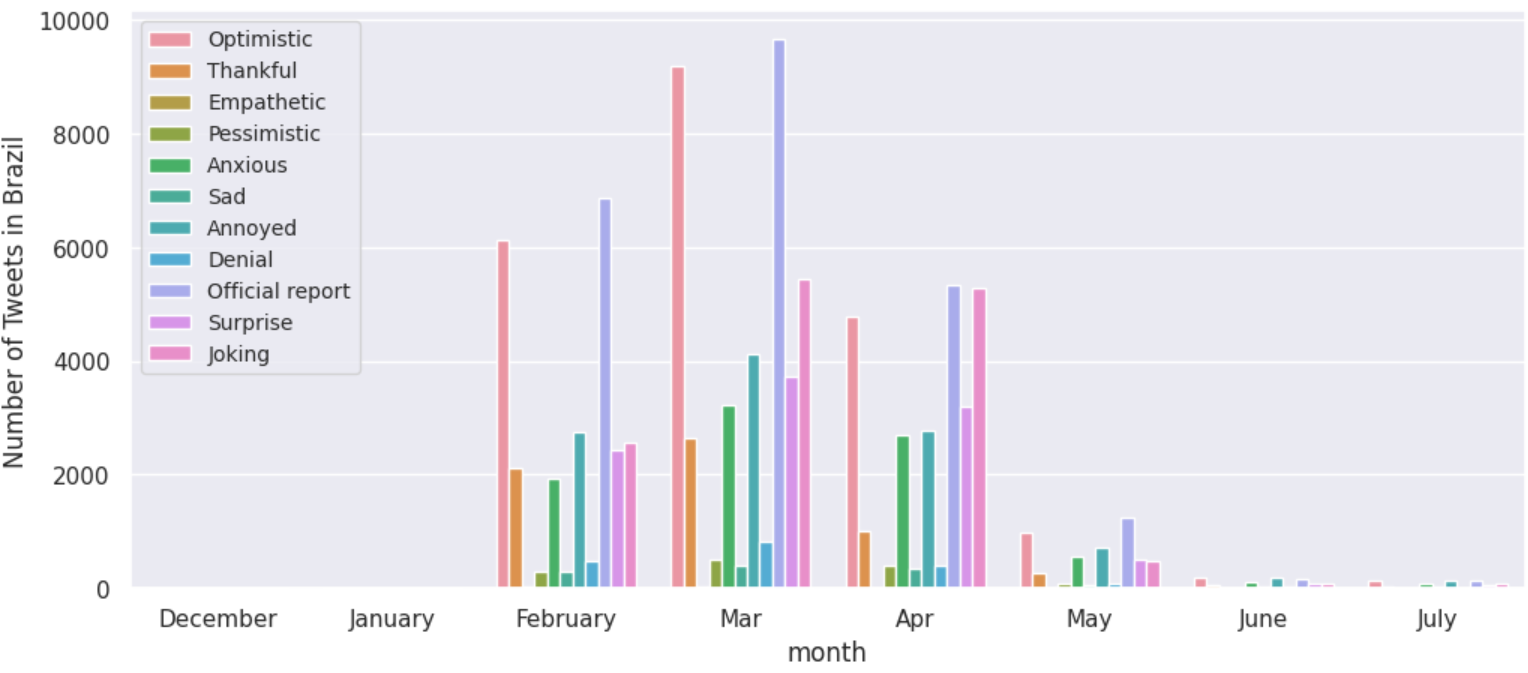}%
}
\subfigure[ August 2021 - January 2022]{
    \includegraphics[height=7cm]{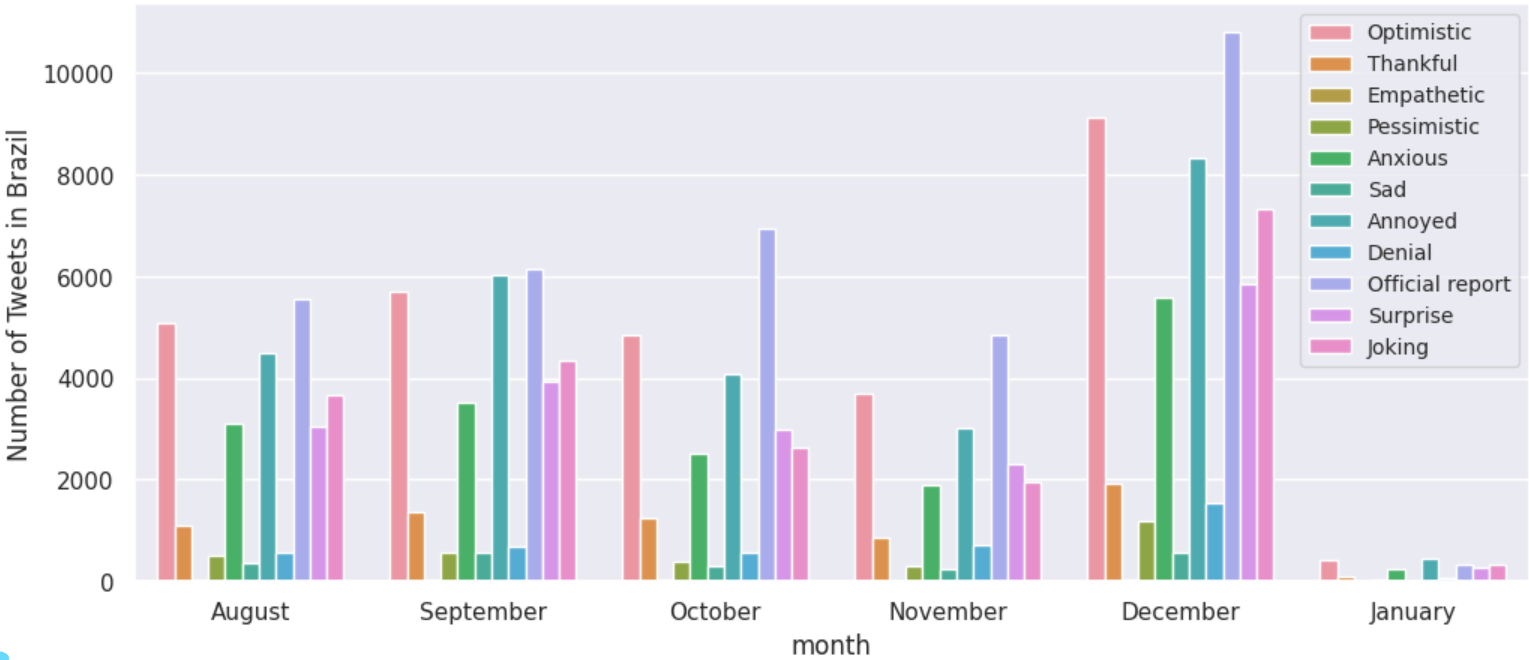}%
}
\caption{ Monthly analysis of the number of tweets for each sentiment for Brazil}
\label{fig:trend} 
\end{figure*}

\begin{figure*}[htbp!]
\centering 
\subfigure[ March 2020 - November 2020]{
    \includegraphics[height=7cm]{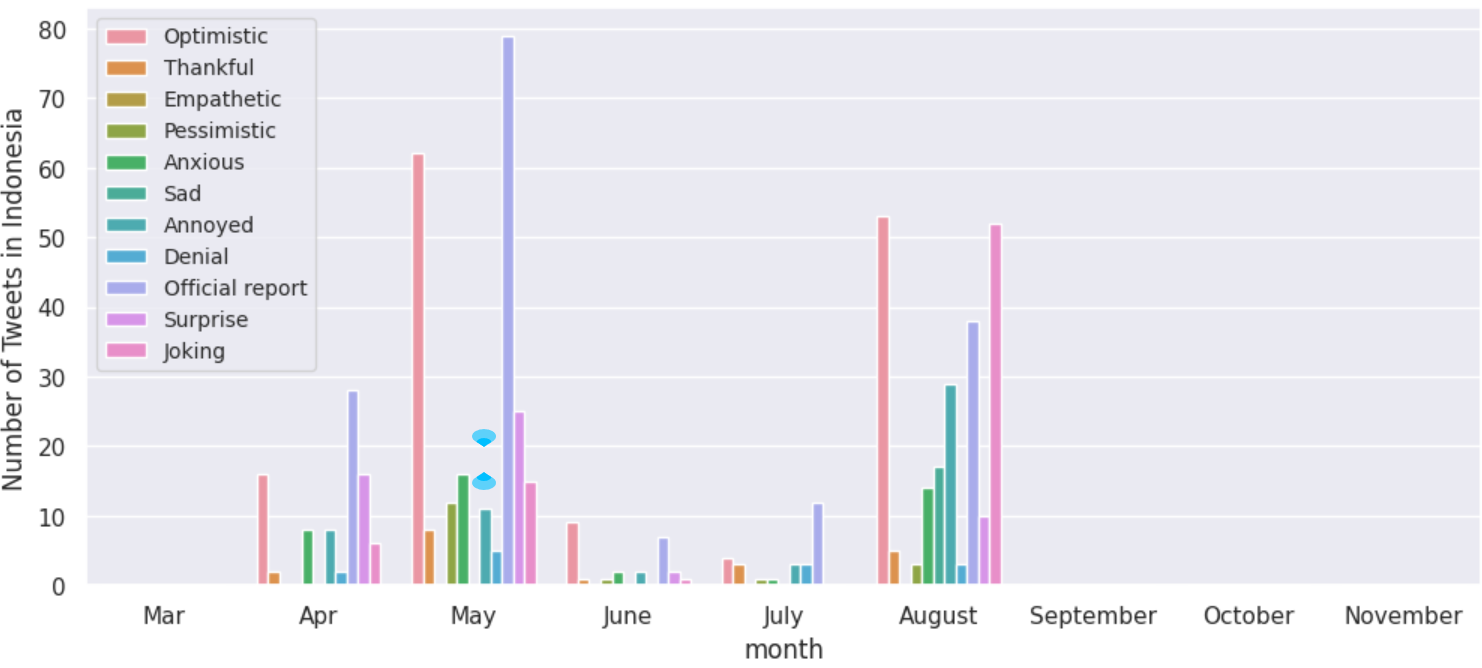}%
}
\subfigure[ December 2020 - July 2021]{
    \includegraphics[height=7cm]{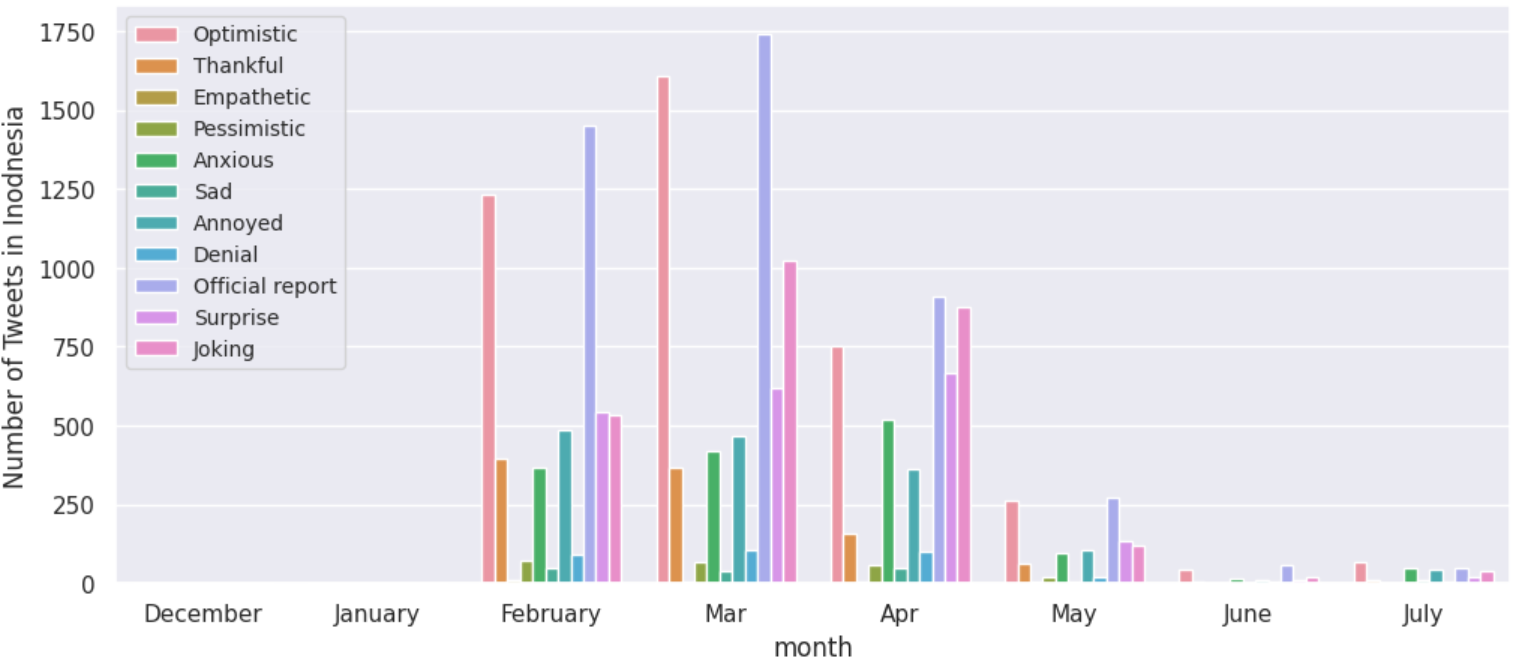}%
}
\subfigure[ August 2021 - January 2022]{
    \includegraphics[height=7cm]{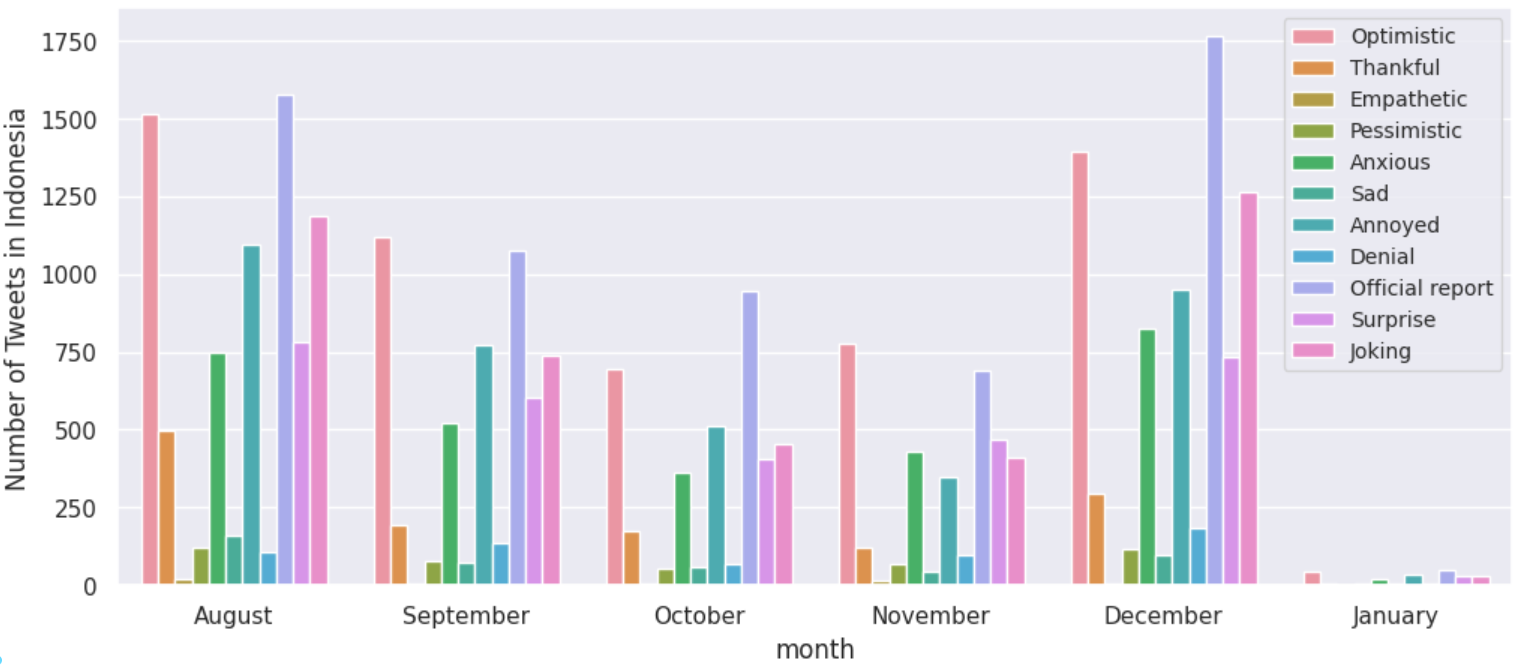}%
}
\caption{ Monthly analysis of the number of tweets for each sentiment for Indonesia}
\label{fig:trend} 
\end{figure*}

\end{document}